\documentclass[a4paper]{IEEEtran}

\usepackage{enumerate}
\usepackage{multirow,array}
\usepackage{cite}
\usepackage{graphicx}
\usepackage{psfrag}
\usepackage{caption}
\usepackage{subcaption}
\usepackage{url}
\usepackage{amsmath}
\usepackage{amsthm}
\usepackage{array}
\usepackage{amssymb}
\usepackage{amsfonts}
\usepackage{float}
\usepackage{tabu}
\usepackage{algorithm}
\usepackage{algorithmicx}
\usepackage{algcompatible}
\usepackage{algpseudocode}

\usepackage{color, colortbl}
\definecolor{Gray}{gray}{0.9}

\makeatletter
\renewcommand{\boxed}[1]{\text{\fboxsep=.2em\fbox{\m@th$\displaystyle#1$}}}
\makeatother
\newcommand{\C}{\cal}

\newcounter{casenum}
\newenvironment{caseof}{\setcounter{casenum}{1}}{\vskip.5\baselineskip}
\newcommand{\case}[2]{\vskip.5\baselineskip\par\noindent {\bfseries Case \arabic{casenum}:} #1\\#2\addtocounter{casenum}{1}}
\newtheorem*{conjecture*}{Conjecture}
\newtheorem{lemma}{Lemma}

\newtheorem{definition}{Definition}
\newtheorem{theorem}{Theorem}
\newtheorem{example}{Example}
\newtheorem{claim}{Claim}

\title{Coded Data Rebalancing for Decentralized Distributed Databases}
\begin{document}

\author{
\IEEEauthorblockN{K V Sushena Sree, Prasad Krishnan\\}
%\IEEEauthorblockA{
%Signal Processing and Communications Research Center,\\
%International Institute of Information Technology, Hyderabad, India.\\
%Email: sushenasree@gmail.com, prasad.krishnan@iiit.ac.in}
% \vspace{-0.5cm}
}

\date{\today}
\maketitle
\thispagestyle{empty}	
\pagestyle{empty}
%%%%%%%%
\begin{abstract}
The performance of replication-based distributed databases is affected due to non-uniform storage across storage nodes  (also called \textit{data skew}) and reduction in the replication factor during operation, particularly due to node additions or removals. Data rebalancing refers to the communication involved between the nodes in correcting this data skew, while maintaining the replication factor. For carefully designed distributed databases, transmitting coded symbols during the rebalancing phase has been recently shown to reduce the communication load of rebalancing. In this work, we look at balanced distributed databases with \textit{random placement}, in which each data segment is stored in a random subset of $r$ nodes in the system, where $r$ refers to the replication factor of the distributed database. We call these as decentralized databases. For a natural class of such decentralized databases, we propose rebalancing schemes for correcting data skew and the reduction in the replication factor arising due to a single node addition or removal. We give converse arguments which show that our proposed rebalancing schemes are optimal asymptotically in the size of the file. 
\end{abstract}
%%%%
\let\thefootnote\relax\footnotetext{
 Sushena Sree and Dr.\ Krishnan are with the Signal Processing \& Communications Research Center, International Institute of Information Technology Hyderabad, India, email:sushenasree@gmail.com, prasad.krishnan@iiit.ac.in.
}

% \begin{IEEEkeywords}
% coded caching, interference management, low subpacketization, projective geometry.
% \end{IEEEkeywords}

\section{Introduction}
Large scale data storage as well as data analytics engines crucially rely upon reliable distributed database systems to efficiently store and process data. The imbalance in distribution of data across the storage nodes is one of the prime factors due to which data stores and analytics platforms are found to underperform. This imbalance is termed as data skew \cite{WhyDataSkew}.  In order to rectify data-skew, most distributed databases or file systems employ a simple technique called \textit{data rebalancing} \cite{ApacheIgniteDataRebalancing,ApacheHadoopDataRebalancing,GoogleCloud,CephRebalancing,AdaptiveDataPlacementforParallel(REBALANCINGOLDPAPER)}. In data rebalancing, the data is moved between the storage nodes so that all nodes store approximately same amount of data, thus reducing data skew. Further, if the database has the data replicated with some replication factor, the rebalancing scheme has to ensure that this replication factor is not reduced during rebalancing. Efficient data rebalancing algorithms are those in which the communication involved during the rebalancing is kept minimal. 

Data rebalancing was formally introduced and studied in \cite{codedData} (coauthored by a subset of the present authors). In \cite{codedData}, data rebalancing schemes were presented for correcting the data skew and replication factor reduction caused by single node removal and addition. A matching converse was also presented in \cite{codedData}, hence showing that these rebalancing schemes have optimal communication loads. The initial distributed database for which these rebalancing schemes were constructed in \cite{codedData} were known as \textit{$r$-balanced distributed databases}, and were designed carefully with a specific structure which will ensure that the communication load due to rebalancing are minimum. Thus, these databases must be centrally designed by some coordinator node and the data must be placed in the storage nodes according to this design. However, such central design may not always be possible in all scenarios. For instance, when new data arrives at some intervals to be stored in the system, a central design of the database may not be feasible. This motivates a flexible \textit{decentralized} design, in which each data segment can be stored in some random subset of nodes in the database independently of other segments. 

In this paper, we consider design rebalancing schemes for a natural class of decentralized distributed databases. These decentralized distributed databases are \textit{$r$-balanced}, i.e., the replication factor for each data segment in the database is $r$, and the expected number of bits stored in each node is the same. 

For such decentralized $r$-balanced databases, we present rebalancing schemes for single node addition and removal scenarios. The rebalancing schemes ensure that both the replication factor and the balanced property of the decentralized database is maintained. We also present information theoretic lower bounds on the expected communication load, and thus show that our rebalancing schemes are optimal asymptotically in the size of the data. Further, this asymptotic communication load is equal to the optimal communication load for rebalancing in centralized databases as shown in \cite{codedData}.

The paper is organized as follows. Section \ref{sys_model} describes the system model and the definition of a decentralized $r$-balanced distributed database, giving a natural construction for the same. Formal definitions of the rebalancing schemes and their associated expected communication loads is also described in this section. The coded data rebalancing scheme for node removal and node addition is described in Section \ref{section:nodeRemovalRebalancing} and Section \ref{section:nodeAdditionRebalancing} respectively. The communication load in each case along with the respective converses are also presented in these sections.

\textit{Related work:}  Decentralized data storage designs have been considered in literature in the context of erasure coded distributed storage, for instance in \cite{DecentralizedErasureCodes}. In \cite{DecentralizedErasureCodes}, $k$ symbols of the data are encoded via a random generator matrix and the encoded segments are stored in $n$ nodes, and the recovery properties of this decentralized erasure code is studied. A similar decentralized distributed encoding structure was explored using fountain codes \cite{DataPersistenceDecentralizedDigitalFountain} in the context of wireless sensor networks. Our work however considers replication-based decentralized storage, with focus on the rebalancing problem rather than the data recovery problem.

The idea of exploiting local storage to reduce communication load by coding together symbols demanded by multiple nodes, is well explored in recent literature, especially in coded caching \cite{MaN} and distributed computing \cite{FundLimitsDistribCom}. Our rebalancing schemes are also related to the transmission schemes of such works in this sense. Earlier work in coded caching \cite{decentralizedcodedcaching} also considers decentralized data placement; however what this means in \cite{decentralizedcodedcaching} is that each client node independently caches some fraction of each file in the file library, chosen randomly. Our decentralized database structure however differs from this, as it refers to placement of each data segment independently in some random subset of nodes. Further, none of these existing works focus on data \textit{rebalancing}, while this is the chief focus of our present work in the context of our decentralized databases. 

%%%%
\textit{Notations and Terminology:} $\mathbb{Z}^{+}$ denotes the set of positive integers. We denote the set $\{1,\hdots,n\}$ by $[n]$ f
or some $n \in \mathbb{Z}^{+}$. For sets $A,B$, the set of elements in $A$ but not in $B$ is denoted by $A\backslash B$.  For a set ${\cal X}$ and some positive integer $t\leq {|\cal X|}$, we denote the set of all $t$-sized subsets of ${\cal X}$ by $\binom{{\cal X}}{t}$. The union of some set $A$ with an element $k$ is denoted by $A\cup k.$ The binomial distribution with parameters $n$ and $p$ is given as $B(n,p).$

%Let $\mathbb{P}(A)$ and $\mathbb{I}(A)$ denote the probability and the indicator function of an event $A$ respectively.
%$\mathbb{E}(X)$ denotes the expectation of random variable $X$.
%$|.|$ denotes the cardinality of a set. 

%%%%%%
\section{System Model}
\label{sys_model}
%%%%%%%%%
Consider a file $W$ consisting of a set of $F$ segments where the $i^{th}$ segment is denoted as $W_i$ for $i \in [F]$. Without loss of generality, we consider the $W_i$s as bits. The system consists of $K$ nodes indexed by $[K]$. Each node $k \in [K]$ is connected to every other node $[K]\setminus k$ via a bus link. This facilitates a noise-free broadcast channel between the nodes. In \cite{codedData}, the idea of distributed database was defined as follows:
%%%%%%%%%%%%%%%%%%%%
\begin{definition}[Distributed Database and Replication factor]
\label{distributedDatabase}
A \textit{distributed database} of $W$  across the nodes $[K]$ consists of a collection $\cal{D}$ of subsets of $W$,
\begin{align*}
    {\cal{D}}=\{D_n \subseteq W: n \in [K]\},
\end{align*}
such that $\bigcup\limits_{n \in [K]} D_n=W$, where $D_n$ denotes the set of bits stored at node $n$. 
%%%%%%%%%%%%%%%%%%%%%%
% W_i \in W_{\overline{S}}, n \notin \overline{S}
Given a distributed database $\cal{D}$ and a subset of nodes $S \subset [K]$, the replication factor of bit $W_i$, denoted by $r_i$ is defined as the number of nodes in which $W_i$ is stored. 
\end{definition}

We assume that the file $W$ is distributed across the $K$ nodes under some random placement strategy such that each bit is stored in the nodes (in at least one node) independently according to some probability distribution. We thus obtain a distributed database with some replication factor for each bit. We call this as a \textit{decentralized distributed database}. In this work, we consider a class of decentralized databases that we define below.

\begin{definition}[Decentralized $r$-balanced distributed database]
\label{defn r bal}
A decentralized $r$-balanced database on $K$ nodes is a \textit{distributed database} denoted by ${\cal{D}}(r,[K])=\{D_n \subseteq W: n \in [K]\}$ constructed by random placement such that,
\begin{enumerate}[i)]
    \item \underline{Replication factor condition}: The replication factor of each bit is $r$,
\begin{align*}
    r_i=r, \forall i \in [F],
\end{align*}
    \item \underline{Balanced state condition}: The expected number of bits stored in each node is same. As the number of bits in the nodes is $rF,$ this means, for each $n\in[K]$, we must have
\[\mathbb{E}(|D_n|)=\lambda F \text{ where } \lambda \triangleq \frac{r}{K}
\text{ is the storage fraction.}\]
\end{enumerate}
\end{definition}

Let $S \subset [K]$ denote a set of nodes. The collection of bits that are exclusively stored at $S$ and thus not available at $[K]\setminus S$ is denoted by $W_{\overline{S}}$ where $\overline{S} = [K] \setminus S$. 
Let $N_i$ denote the set of nodes where the bit $W_i$ is stored during initial storage placement. Thus, the event $(N_i=S)$ indicates that $W_i \in W_{\overline{S}}$. Further, the event that a bit $W_i \in W$ is stored at node $n \in [K]$ implies that $W_i \in W_{\overline{S}}:n \notin \overline{S}$.

%(W_i \in W_{\overline{S}}: \overline{S}=[K] \setminus S)
%For $r \in {\mathbb{Z}^{+}}: r>1$, we define a decentralized $r$-balanced database denoted by ${\cal{D}}(r,[K])$ of a file $W$ of size $F$ bits distributed randomly according to some probability distribution across nodes $[K]$, as a collection $\{D_k \subseteq W: k \in [K]\}$ such that 

%(ii) $\mathbb{P}(W_i \in W_{\boldsymbol{m}}: \boldsymbol{m} \in \binom{[K]}{K-r})=\frac{1}{\binom{K}{r}}$
%(ii) $\mathbb{P}(N_i \in \binom{[K]}{r})=\frac{1}{\binom{K}{r}}$
%%%%%%%%%%%%%%%%%%%%%%%%%%%%%%%%%%%%

In a decentralized setup, there is no central node to coordinate the storage placement of the bits across the nodes. Every bit is independently stored in the system.  The following lemma describes a natural method to create a decentralized $r$-balanced distributed database of the file $W$.  We shall also use this lemma to check the $r$-balanced property after rebalancing. 
%%%%%%%%%%%%%%%%%%%%%%%%%%So, we allow each bit to choose the nodes where it will be stored.
\begin{lemma}
\label{uniform}
Consider a distributed database created as follows: 
\begin{itemize}
    \item Each bit is stored in a set of $r$ nodes chosen independently and uniformly at random from the set of $K$ nodes (i.e)
\end{itemize}
%When each bit $W_i:i \in [F]$ is allowed to choose randomly and independently a set of nodes $S: S \subset [K]$ where it will be stored such that \mathbb{P}(N_i=S)=
    % \begin{cases}
    % \frac{1}{\binom{K}{r}}, & \text{if } S \in \binom{[K]}{r}\\
    % 0, & otherwise.
    % \end{cases}
\begin{align*}
    \mathbb{P}(N_i=S)=
    \frac{1}{\binom{K}{r}},~~~\forall S \in \binom{[K]}{r}
\end{align*}
Then the resultant database is a decentralized $r$-balanced distributed database. 
\end{lemma}
%%%%%%%%%%%%%%%%%%%%%%%%%%%%%%%To obtain a decentralized $r$-balanced distributed database the conditions specified in Definition \ref{defn r bal} must be satisfied.
\begin{IEEEproof}
We check whether the conditions in Definition \ref{defn r bal} are satisfied. Following the lemma statement, by allowing each bit $W_i$ to be stored exclusively at a set of nodes $S: S\in \binom{[K]}{r}$, it is easy to see that $r_i=r, \forall i \in [F]$. This satisfies the replication factor condition of Definition \ref{defn r bal}. 

With the above assignment strategy, for each $n \in [K]$ and $i \in [F]$, the probability that $W_i$ is stored at a node $n \in [K]$ is given by,
\begin{align*}
    \mathbb{P}(W_i \in D_n) &= \mathbb{P}\Bigg( \bigcup\limits_{S \in \binom{[K]}{r}: n \in S} (N_i = S) \Bigg) \\
    &= \sum\limits_{S \in \binom{[K]}{r}: n \in S} \mathbb{P}(N_i=S)\\
    &= \frac{\binom{K-1}{r-1}}{\binom{K}{r}}\\
    &= \frac{r}{K} = \lambda.\\
\end{align*}
The expected number of bits stored at node $n \in [K]$ is calculated as,
\begin{align*}
{\mathbb{E}}(|D_n|)&={\mathbb{E}}(\sum_{i \in [F]}{\mathbb{I}}(W_i \in D_n)=\sum_{i \in [F]}\lambda=\lambda F. 
\end{align*}
%\begin{align}
%\label{exp}
%    {\mathbb{E}}(|D_k|)={\mathbb{E}}(\sum_{i \in [F]}{\mathbb{I}}(W_i \in W_{[K]\setminus S }: k \in S))=\sum_{i \in [F]}\lambda=\lambda F. 
%\end{align}
This satisfies the balanced state condition of Definition \ref{defn r bal}, which completes the proof. 
\end{IEEEproof}
%%%Hence the decentralized random placement of bits across the nodes with constraint on the number of nodes where each bit is stored as $r$ ensures that we obtain a decentralized $r$-balanced distributed database.
%We now describe how to create a decentralized $r$-balanced database of the file $W$. In a decentralized setup, there is no central server to coordinate the cache placement of the bits across the nodes. We allow each bit $W_i:i \in [F]$ to choose $r$ nodes out of the $K$ nodes in the system randomly and independently with uniform probability. We see that $r_i([K])=r, \forall i \in [F]$. Let $S \in \binom{[K]}{r}$ be a set of $r$ nodes chosen by $W_i$ and let $k \in [K]$. Assuming uniform probability over the set $\{S: S \in \binom{[K]}{r}\}$,  

%$\mathbb{P}(W_i \in W_{[K]\setminus S})=\frac{1}{\binom{K}{r}}$. Consider a node $k \in [K]$,

When a node is removed (or added) to the system, the replication factor condition and the balanced state condition is disrupted. To restore the decentralized $r$-balanced distributed database, rebalancing operation involving transmission of bits among the nodes is necessary. We next formally describe the rebalancing strategies in case of node removal and node addition and also the expected communication load associated with each case.

\subsection{Node Removal}
%When a node is removed or added to a system, it causes imbalance to $r$-balanced database in terms of the replication factor. This necessitates rebalancing operation through transmissions among the nodes to restore the replication factor. We next formally describe the rebalancing strategies in case of node removal and node addition and also the communication load involved in each case.
%%%%%%%%%%%%%%%%%%%%%%%%%%%%%%%%%%%%%5
Given a decentralized $r$-balanced distributed database ${\cal{D}}(r,[K])$, let us consider a scenario where a node $k \in [K]$ is removed. Let ${\cal{D}}_k(r,[K]\setminus k)= \{D_n^k: n \in [K] \setminus k\}$ be the target decentralized $r$-balanced distributed database that we want to accomplish after rebalancing operation in the updated system consisting of nodes $[K] \setminus k$. 

\begin{figure}
    \centering
    \includegraphics[width=\linewidth,trim={ 0 0 500 0},clip]{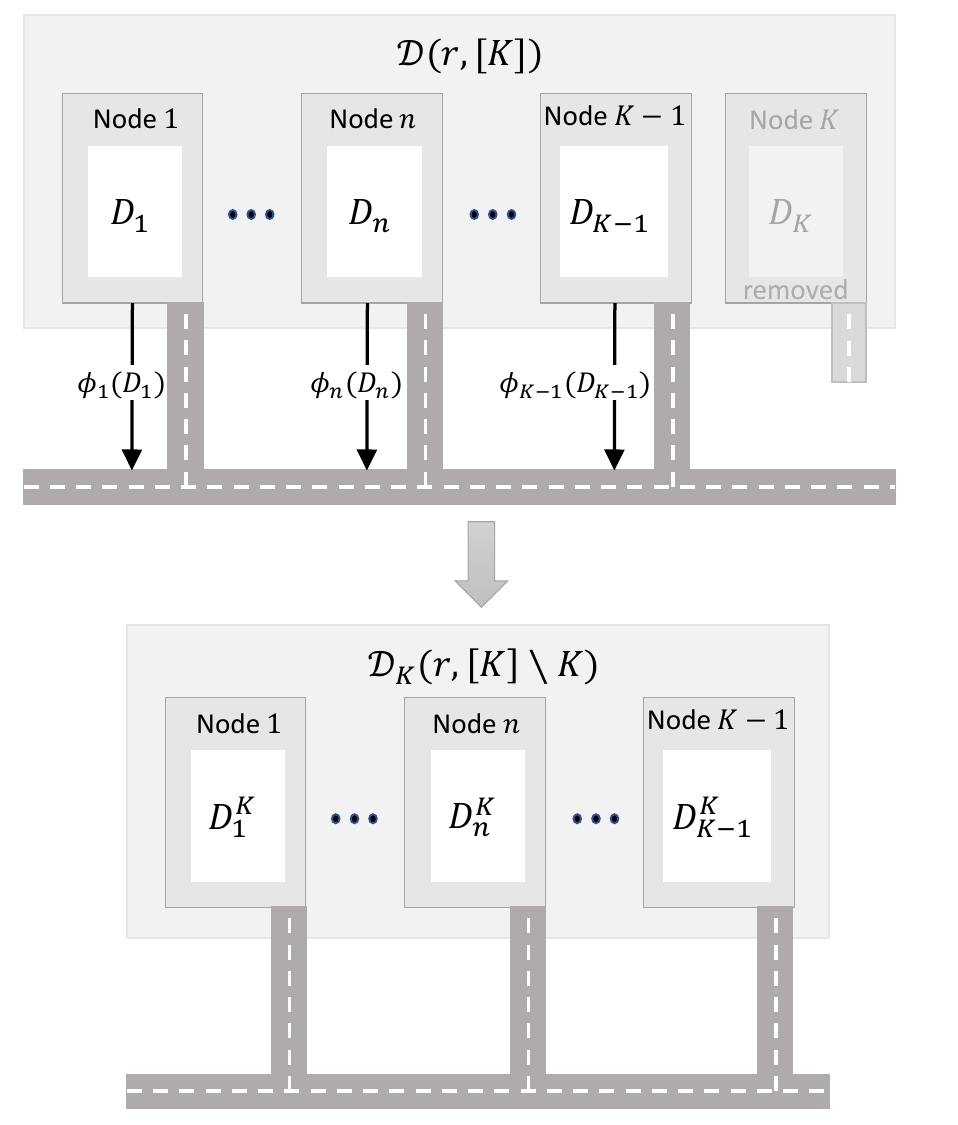}
    \caption{Rebalancing scheme ${\cal{R}}(K,{\cal{D}},{\cal{D}}_K)$ after removal of node $K$ from the database ${\cal{D}}(r,[K])$. Each surviving node $n$ broadcasts codeword $\phi_n(D_n)$. The storage content at each node $n$ is then updated to $D_n^K$ using its previous storage content $D_n$ and the $K-2$ transmissions from the remaining nodes to obtain the target database ${\cal{D}}_K(r,[K] \setminus K).$}
    \label{fig:nodeRemoval}
\end{figure}

Generally, a rebalancing scheme for node removal denoted by ${\cal{R}}(k,{\cal{D}},{\cal{D}}_k)$ comprises of a collection of encoding functions $\{\phi_n:\forall n\in[K]\backslash k\}$ and decoding functions $\{\psi_n:\forall n\in[K]\backslash k\}$. From each node $n\in[K]\backslash k$, a codeword $\phi_n(D_n)$ of length $l_n$ bits is broadcasted to all the remaining surviving nodes. Each surviving node $n \neq k$ should be able to decode its demand $D_n^k$ by applying the decoding function $\psi_n$ over the current storage content $D_n$ and the received codewords from other surviving nodes. Figure \ref{fig:nodeRemoval} illustrates this process.

% \begin{definition}
% (Rebalancing scheme for node removal) Let $l_n \in \mathbb{Z}^+: i \in [K] \setminus k$. We define a \underline{rebalancing scheme} from ${\cal{D}}(r,[K])$ to ${\cal{D}}_k(r,[K]\setminus k)$ by ${\cal{R}}(k,{\cal{D}},{\cal{D}}_k) \triangleq \{ \phi_n, \psi_n : n \in [K] \setminus k\}$ which comprises of a set of encoding functions,
% \[\phi_n: \{0,1\}^{|D_n|} \rightarrow \{0,1\}^{l_n}, for~ each~ n \in [K] \setminus k,\]
% and a set of decoding functions,
% \[\psi_n:\{0,1\}^{|D_n|} \times \prod\limits_{j \in [K] \setminus \{n,k\}} \{0,1\}^{l_j} \rightarrow \{0,1\}^{|D_n^k|},\]
% for each $n \in [K] \setminus k$ such that 
% \[\psi_n(D_n,(\phi_j(D_j): j \in [K] \setminus \{n,k\}))=D_n^k.\]
% \end{definition}
The \underline{expected communication load} of the coded data rebalancing scheme for node removal ${\cal{R}}(k,{\cal{D}},{\cal{D}}_k)$, is given by the expected number of transmitted bits normalized by the expected number of bits stored in the removed node $k$ (i.e) $\mathbb{E}(|D_k|)$, which is denoted by 
\begin{align*}
    C_{rem}({\cal{R}}(k,{\cal{D}},{\cal{D}}_k))\triangleq \frac{\mathbb{E}\Big(\sum\limits_{n \in [K] \setminus k}l_n\Big)}{\lambda F}.
\end{align*}
% The \underline{expected asymptotic communication load} of the coded data rebalancing scheme for node removal is obtained by calculating the expected communication load under the condition of large number of bits present in the database which is given by,
% \begin{align*}
%     C_{rem}({\cal{R}}(k,F,{\cal{D}},{\cal{D}}_k)) \triangleq \underset{F \rightarrow \infty}{lim}C_{rem}({\cal{R}}(k,{\cal{D}},{\cal{D}}_k))
% \end{align*}
%%%%%%%%%%%%%%%%%%%%%%%%%
The optimal rebalancing load under node removal for replication factor $r$ is given as 
\begin{align*}
    C_{rem}^*(r) = \inf  ~C_{rem}({\cal{R}}(k,{\cal{D}},{\cal{D}}_k)),
\end{align*}
where the infimum is taken over all possible choices for (a) the initial $r$-balanced decentralized database ${\cal D}$ (b) the collection of $r$-balanced target databases $\{{\cal D}_k:k\in[K]\}$, and (c) the rebalancing schemes, given by $\{{\cal{R}}(k,{\cal{D}},{\cal{D}}_k):k\in[K]\}$. 

\subsection{Node Addition}
Consider a new node indexed by $K+1$ added to the system of nodes $[K]$. The new node is assumed to have no content in its storage during its arrival, and thus a data skew is created in the system. After performing rebalancing operation for node addition we target to achieve a decentralized $r$-balanced distributed database ${\cal{D}}^*(r,[K+1])=\{D_n^* \subset W : n \in [K+1]\}$.  

\begin{figure}
    \centering
    \includegraphics[width=\linewidth,trim={0 0 500 0},clip]{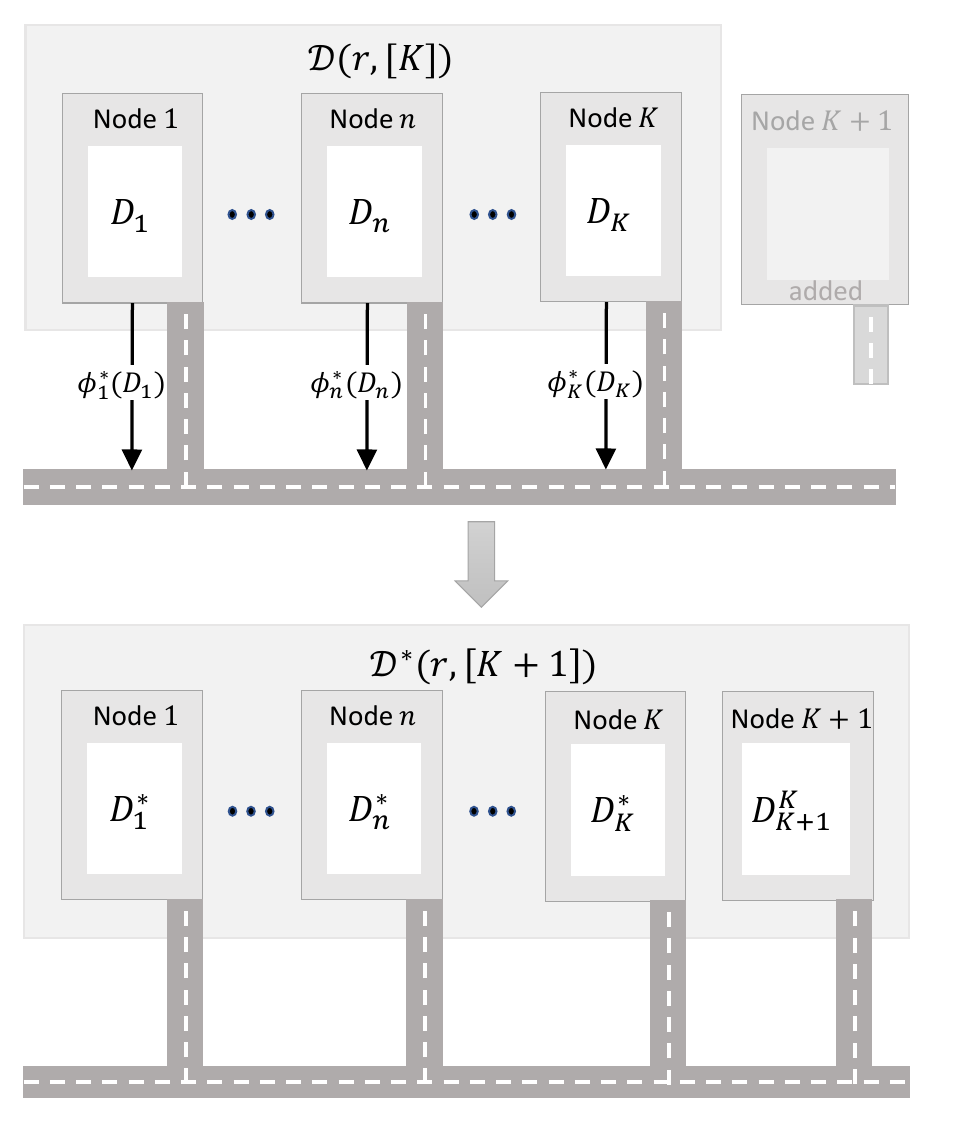}
    \caption{Rebalancing scheme ${\cal{R}}^*({\cal{D}},{\cal{D}}^*)$ after addition of node $K+1$ to the database ${\cal{D}}(r,[K])$. Each pre-existing node $n$ broadcasts codeword $\phi^*(D_n)$. The new node's storage is updated to $D_{K+1}^*$ using these transmissions. The storage content at each pre-existing node $n$ is then modified to $D_n^*$ to obtain the target database ${\cal{D}}^*(r,[K+1]).$}
    \label{fig:nodeAddition}
\end{figure}

In general, a rebalancing scheme for node addition consists of a collection of encoding functions $\{\phi_n:\forall n\in[K]\}$ and decoding functions $\{\psi_n:\forall n\in[K+1]\}$. Each pre-existing node $n \in [K]$ broadcasts a codeword $\phi_n^*(D_n)$ of length $l_n$. Using the received codewords, the new node decodes using a decoding function $\psi_{K+1}^*(\phi_n^*(D_n):n \in [K])=D_{K+1}^*$. Each pre-existing node $n \in [K]$ decodes its demand $D_n^*$ by applying its own decoding function as $\psi_{n}^*(D_n,(\phi_j^*(D_j): j \in [K] \setminus n))= D_n^*$. The process is illustrated in Figure \ref{fig:nodeAddition}.
% \begin{definition}
% (Rebalancing scheme for node addition) 
% Let $l_i \in \mathbb{Z}^+ : i \in [K] \setminus k$. We define a rebalancing scheme from ${\cal{D}}(r,[K])$ to ${\cal{D}}^*(r,[K+1])$ by ${\cal{R}}^*({\cal{D}},{\cal{D}}^*) \triangleq \{ \phi_n^*, \psi_j^* : n \in [K], j \in [K+1]\}$ which comprises of a set of encoding functions, $\phi_n^*: \{0,1\}^{|D_n|} \rightarrow \{0,1\}^{l_n}$, for each $n \in [K]$ and a set of decoding functions 
%Let $l_i \in \mathbb{Z}^+: i \in [K] \setminus k$. We define a rebalancing scheme from ${\cal{D}}(r,[K])$ to ${\cal{D}}^*(r,[K+1])$ by ${\cal{R}}^*({\cal{D}},{\cal{D}}^*) \triangleq \{ \phi_n^*, \psi_j^* : n \in [K], j \in [K+1]\}$ which comprises of a set of encoding functions,
% \[\phi_n^*: \{0,1\}^{|D_n|} \rightarrow \{0,1\}^{l_n}, for~ each~ n \in [K] \]
% and a set of decoding functions,
% \begin{itemize}
%     \item $\psi_n^*:\{0,1\}^{|D_n|} \times \prod\limits_{j \in [K] \setminus n} \{0,1\}^{l_j} \rightarrow \{0,1\}^{|D_n^*|}$,
% for each $n \in [K]$ such that 
% \[\psi_n^*(D_n,(\phi_j^*(D_j): j \in [K] \setminus \{n\}))=D_n^*, \forall n \in [K]\]
%     \item $\psi_{K+1}^*: \prod\limits_{j \in [K]} \{0,1\}^{l_j} \rightarrow \{0,1\}^{|D_{K+1}^*|}$, such that 
% \[\psi_{K+1}^*(\phi_j^*(D_j): j \in [K])=D_{K+1}^*\]
% \end{itemize}

The \underline{expected communication load} of a rebalancing scheme for node addition ${\cal{R}}^*({\cal{D}},{\cal{D}}^*)$, is given by the expected number of transmitted bits normalized by the expected number of bits $|D_{K+1}|$ stored in the new node, which is denoted as
\begin{align*}
    C_{add}({\cal{R}}^*({\cal{D}},{\cal{D}}^*))\triangleq \frac{{\mathbb E}\left(\sum\limits_{n \in [K]}l_n\right)}{\lambda_{add} F}.
\end{align*}
%\end{definition}

The optimal rebalancing load under node addition is given as, 
\begin{align*}
    C_{add}^*(r)= \inf_{{\cal D},{\cal D}^*} \inf_{{\cal{R}}^*({\cal{D}},{\cal{D}}^*)} C_{add}({\cal{R}}^*({\cal{D}},{\cal{D}}^*)).
\end{align*}
%%%%%%%%%%%%%%%%%%%%%%%%The replication factor of every bit stored in node $n$ i.e $r_i(S)=r-1, \forall~ S \in \binom{K}{r}$ such that $n \in S$.

\section{Coded Data Rebalancing for node removal}
\label{section:nodeRemovalRebalancing}
Consider a decentralized $r$-balanced distributed database ${\cal{D}}(r,[K])$ designed as per Lemma \ref{distributedDatabase}. Now we assume node $k \in [K]$ is removed from the system. For every bit $W_i \in D_k$, the replication factor is reduced by $1$. We see that the replication factor condition is not satisfied.  %and there will be no uniform distribution of bits across the nodes.
To restore the compliance with the replication factor and balanced state condition of the  decentralized $r$-balanced distribution database, each collection of bits $W_{\boldsymbol{m}}:\boldsymbol{m} \in \binom{[K]\setminus k}{K-r}$ that was stored in $k$, must be stored at one of the remaining $K-1$ nodes, in such a way that each node finally stores the same expected number of bits. 
%The expected number of bits that were stored in node $k$ before node removal is $\lambda F$. 
By performing the data rebalancing operation after node removal we target to accomplish a decentralized $r$-balanced distributed database ${\cal{D}}_k(r,[K] \setminus k)=\{D_n^k \subseteq W: n \in [K] \setminus k\}$ in the updated system comprising of the nodes $[K] \setminus k$. Let $N_i'$ denote the set of nodes where bit $W_i$ is stored in the new database ${\cal{D}}_k(r,[K] \setminus k)$. We will show that in the target database, we will have
\begin{equation}
\label{equation: probNewDBRem}
    \mathbb{P}(N_i'=S)=\frac{1}{\binom{K-1}{r}}, \forall~i, \forall~ S \in \binom{[K]\setminus k}{r}.
\end{equation}
If our new database satisfies the above condition, then using Lemma \ref{distributedDatabase}, we can show that
%\[\mathbb{E}(|D_n^k|)=\lambda_{rem} F, ~\forall~n \in [K]\setminus k, \]
%where $\lambda_{rem}=\frac{r}{K-1}$. We can see that $\lambda_{rem}$ is the new storage fraction. With this, we see that 
the target database is also a decentralized $r$-balanced distributed database. We show this in Lemma \ref{noderemovalproofofbalanced}. Finally in Theorem \ref{theorem:nodeRemoval}, we obtain an upper bound on the expected communication load of our rebalancing scheme, and show that as $F$ grows large, this load is asymptotically optimal. 

%According to Definition \ref{defn r bal} we require ${\cal{D}}_k(r,[K] \setminus k),$ to have,
%\begin{align*}
%    \mathbb{P}(N_i \in \binom{[K]\setminus k}{r}) &=\mathbb{P}(W_i \in W_{\{[K]\setminus k\} \setminus S}: S \in \binom{[K]\setminus k}{r})\\ &=\frac{1}{\binom{K-1}{r}}
%\end{align*}
%Consider a node $j \in [K]\setminus k,$
%\begin{align*}
%    &\mathbb{P}(N_i \in \binom{[K]\setminus k}{r}: j \in N_i)\\
%    &=\mathbb{P}(W_i \in W_{\{[K]\setminus k\} \setminus S}: S \in \binom{[K]\setminus k}{r},j \in S  )\\
%    &=\frac{\binom{K-2}{r-1}}{\binom{K-1}{r}}=\frac{r}{K-1}\triangleq \lambda_{rem},\\
%    &
%\end{align*}
%Generally the rebalancing scheme for node removal involves the broadcast of a codeword $\phi_i(D_i)$ by each surviving node $i \neq k$ to the remaining surviving nodes. $K-1$ transmissions occur after which each surviving node $i \neq k$ decode their demand $D_i(k)$ by applying the decoding function $\psi_i$ over the current cache content $D_i$ and the received codewords. We illustrate this process in Figure.
%%%%%%%%%%%%%%%%%%%%%%%%%%%
\begin{figure*}
  \includegraphics[width=\textwidth,trim={0 9.8cm 0 0 0},clip]{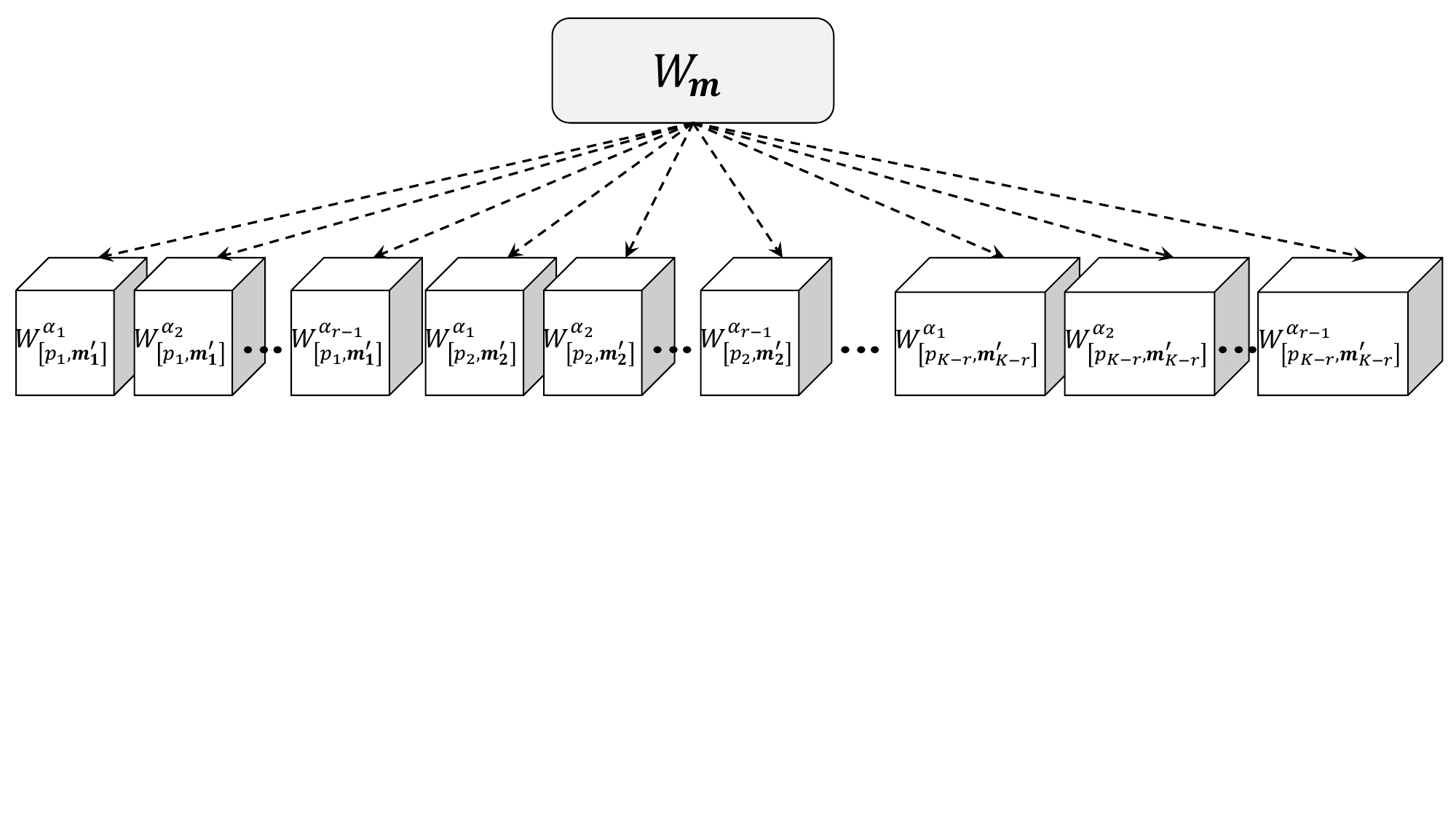}
  \caption{For every $\boldsymbol{m} \in {\cal{A}}_k$, consider $(K-r)(r-1)$ boxes labelled by $W_{[p_j,\boldsymbol{m}_j']}^ \alpha,$ where $p_j \in \boldsymbol{m}, \boldsymbol{m}_j' = \boldsymbol{m} \setminus p_j$ and $\alpha \in [K]\setminus (k \cup \boldsymbol{m})$.
  %$\boldsymbol{m}_j' \in \binom{\boldsymbol{m}}{K-r-1}, p_j \cup \boldsymbol{m}_j'=\boldsymbol{m}$ and $\alpha \in [K]\setminus \{k \cup \boldsymbol{m}\}$. 
  Every bit in $W_{\boldsymbol{m}}$ is associated with one of these boxes chosen uniformly at random.}
  \label{fileSplit}
\end{figure*}
%%%%%%%%%%%%%%%%%%%We now provide a coded data rebalancing scheme to mitigate the data skew caused by the removal of a node $k \in [K]$ from the decentralized $r$-balanced distributed database ${\cal{D}}(r,[K])$ (designed as in Lemma \ref{distributedDatabase})
\subsection{Coded data rebalancing scheme after node removal}
\label{subsection: node removal}
%%%%%%%%%%%%%%%%%%%%%%
We will now elaborate on the rebalancing scheme which is applied on the decentralized $r$-balanced distributed database ${\cal{D}}(r,[K])$ (designed as in Lemma \ref{distributedDatabase}) after removing a node $k \in [K]$. Let us index the collection of bits that were stored in the removed node $k$ by ${\cal{A}}_k$ where \[{\cal{A}}_k= \binom{[K]\setminus k}{K-r}.\]
Recall that for $\boldsymbol{m}\in {\cal A}_k$,  $W_{\boldsymbol{m}}$ refers to the set of bits which are not available in $\boldsymbol{m}$ but available in the $r-1$ survivor nodes $[K]\setminus (\boldsymbol{m} \cup k)$. For each $ \boldsymbol{m} \in {\cal{A}}_k$, consider a set of $(K-r)(r-1)$ boxes that are labelled by $$\{W_{[p_j,\boldsymbol{m}_j']}^ \alpha : \forall p_j \in \boldsymbol{m}, \boldsymbol{m}_j' = \boldsymbol{m} \setminus p_j, \forall \alpha \in [K] \setminus (\boldsymbol{m} \cup k)\}.$$
%\boldsymbol{m}_j' \in \binom{\boldsymbol{m}}{K-r-1}, p_j \cup \boldsymbol{m}_j'=\boldsymbol{m}, \alpha \in [K] \setminus (\boldsymbol{m} \cup k) \}$
We then associate to each bit in the collection of bits $W_{\boldsymbol{m}}$ one box chosen uniformly at random as shown in Figure \ref{fileSplit}. This binning process is performed at some node in $[K]\backslash(\boldsymbol{m}\cup k)$ which contains $W_{\boldsymbol{m}}$, and communicated to all other nodes in $[K]\setminus (\boldsymbol{m} \cup k)$, so that all these nodes have the same bits in the respective bins. We then collectively call the set of bits that have chosen the same box as a    \textit{packet} which is indexed by the label of the box they have chosen in common. For a packet $W_{[p_j,\boldsymbol{m}_j']}^ \alpha$, the survivor node $p_j$ denotes a node where the bits in  $W_{[p_j,\boldsymbol{m}_j']}^ \alpha$ are not stored and $\alpha$ gives the index of a survivor node where the bits are stored.
%$\{p_j \cup \boldsymbol{m}_j'\}$ indexes the set of nodes where the set of bits are stored and $\alpha$ gives the index of a node where the bits are not stored. 
%The number of choices for $p_j$ such that $p_j \notin \binom{[K]\setminus k}{K-r-1}$ and $p_j \in [K]\setminus k$ is $r$. For a particular $\boldsymbol{m}' \in \binom{[K]\setminus k}{K-r-1}$ if we look at the packet $W_{[p_j,\boldsymbol{m}']}^ \alpha$, it is stored at the nodes indexed by $p_1,p_2,...,p_{j-1},p_{j+1},..,p_{r}$ and not stored at the the node indexed by $p_j$. We aim to store the packet $W_{[p_j,\boldsymbol{m}']}^ \alpha$ at the node indexed by $p_j$. We can see that each of the $r$ packets $W_{[p_1,\boldsymbol{m}']}^ {\alpha_1} ,W_{[p_2,\boldsymbol{m}']}^ {\alpha_2},...,W_{[p_r,\boldsymbol{m}']}^ {\alpha_r}$ is stored at a unique $r-1$ sized subset of the $r$ nodes $p_1,p_2,..,p_r$ and is required is required to be stored at the remaining node. To facilitate this we make transmissions according to the procedure described in Algorithm \ref{trans algorithm}.

Consider any $\boldsymbol{m}' \in \binom{[K] \setminus k}{K-r-1}$.
For any such $\boldsymbol{m}'$, consider the set of survivor nodes ${\cal P}_{\boldsymbol{m}'} = \{p_{1},\hdots,p_{r}\}=[K]\setminus (\boldsymbol{m}' \cup  k)$.
For any $p_i \in {\cal P}_{\boldsymbol{m}'}$, consider the set of $r-1$ packets given by $\{ W_{[p_j,\boldsymbol{m}']}^{p_i}: \forall p_j \neq p_i\}$. 
Each packet $W_{[p_j,\boldsymbol{m}']}^{p_i}$, which was available at the removed node $k$, is now available at all survivor nodes $p_l: l \neq j$, but not at node $p_j$. 
We seek to store the bits in this packet precisely in node $p_j$. 
This structure allows these packets to be XORed and transmitted by node $p_i$, provided they have the same size. 
This results in  each node $p_j$ being capable of decoding $ W_{[p_j,\boldsymbol{m}']}^{p_i}$ (as all other packets in the XOR are available at $p_j$). 
The algorithm describing the complete rebalancing is shown in Algorithm \ref{trans algorithm}. 

\begin{algorithm}
\caption{Coded data rebalancing transmission scheme for node removal}
\label{trans algorithm}
\begin{algorithmic}[1]
\Procedure{Transmission}{}
    \For{each $\boldsymbol{m}' \in \binom{[K]\setminus k}{K-r-1}$}
    \State Let $\{p_{1},\hdots,p_{r}\}=[K]\setminus (\boldsymbol{m}' \cup  k)$
    \For {each $p_i \in [K]\setminus (k \cup \boldsymbol{m}')$}
    \State {Pad each packet $W_{[p_l,\boldsymbol{m}']}^{p_i}: p_l \neq p_i$ with}
    \Statex {\hspace{\algorithmicindent}
    \hspace{\algorithmicindent}
    \hspace{\algorithmicindent}
    dummy zero bits such that,} \Statex{\hspace{\algorithmicindent} \hspace{\algorithmicindent} \hspace{\algorithmicindent} $|W_{[p_l,\boldsymbol{m}']}^{p_i}|=\max\{|W_{[p_l,\boldsymbol{m}']}^{p_i}|: p_l \neq p_i\}$}
%     \hspace{\algorithmicindent} \hspace{\algorithmicindent} \hspace{\algorithmicindent} \hspace{\algorithmicindent} $,|W_{[p_{i-1},\boldsymbol{m}']}^{p_i}|,|W_{[p_{i+1},\boldsymbol{m}']}^{p_i}|,...,|W_{[p_r,\boldsymbol{m}']}^{p_i}|\}.$
    \State Node $p_i$ transmits $X_{p_i,\boldsymbol{m}'}= \bigoplus\limits_{p_l \neq p_i} W_{[p_l,\boldsymbol{m}']}^{p_i}$
    \EndFor
    \EndFor
\EndProcedure
\end{algorithmic}
\end{algorithm}

After the transmission procedure, each node indexed by $p_j$ decodes its demand $W_{[p_j,\boldsymbol{m}']}^{p_i}$ from the transmission $X_{p_i,\boldsymbol{m}'}$ and its storage content as follows:

\begin{align*}
    &X_{p_i,\boldsymbol{m}'} \oplus \bigg( \bigoplus\limits_{p_l \neq p_j, p_i} W_{[p_l,\boldsymbol{m}']}^{p_i} \bigg)\\
    &= \bigg( \bigoplus\limits_{p_l \neq p_i} W_{[p_l,\boldsymbol{m}']}^{p_i} \bigg) \oplus \bigg( \bigoplus\limits_{p_l \neq p_j, p_i} W_{[p_l,\boldsymbol{m}']}^{p_i} \bigg)\\
    &= W_{[p_j,\boldsymbol{m}']}^{p_i}
\end{align*}
Thus each demanded packet $W_{[p_j,\boldsymbol{m}']}^{p_i}$ is  decoded and stored at node $p_j$ precisely. Once the algorithm is complete, we refer to the resultant distributed database as ${\cal D}_k(r,[K]\backslash k)$. In the next lemma, we show that the obtained distibuted database ${\cal D}_k(r,[K]\backslash k)$ is a decentralized $r$-balanced distributed database.
\begin{lemma}
\label{noderemovalproofofbalanced}
The database ${\cal{D}}_k(r,[K]\setminus k)$ is a decentralized $r$-balanced distributed database. 
\end{lemma}

\begin{IEEEproof}
We first note that any  bit in removed node $k$ is present in the collection $W_{\boldsymbol{m}}$ for some $\boldsymbol{m} \in {\cal A}_k$. By the splitting process described in the Section \ref{subsection: node removal} and Algorithm \ref{trans algorithm}, each bit in $W_{\boldsymbol{m}}$ necessarily appears in some $W_{[p_j,\boldsymbol{m}']}^{p_l}$ for some $p_j,p_l,\boldsymbol{m}'$ such that $p_j\cup \boldsymbol{m}' = \boldsymbol{m}$. By the verification of decoding, every bit in each $W_{\boldsymbol{m}}$ is delivered to a node where it was previously unavailable. Hence the replication factor is reinstated to be $r$ for the bits in node $k$. 
%Consider a packet $W_{[p_j,\boldsymbol{m}']}^{p_i}$ which will be a part of the transmission $X_{p_i,\boldsymbol{m}'}$ sent by node $p_i$ as in Algorithm \ref{trans algorithm}. We can easily see that $p_j$ is the only node in the system that contains packets $\{W_{[p_l,\boldsymbol{m}']}^{p_i} : p_l \neq p_j,p_i$\} in its storage with which it can decode $W_{[p_j,\boldsymbol{m}']}^{p_i}$ from the transmission $X_{p_i,\boldsymbol{m}'}$. We can see that the replication factor of every bit that was previously stored in the removed node $k$, gets updated to $r$. 
The replication factor of the bits that were not initially stored in $k$ is $r$ and it remains unaltered by the rebalancing scheme. Thus the replication factor of every bit in the new database ${\cal{D}}_k(r,[K]\setminus k)$ is $r$. Therefore the replication factor condition of Definition \ref{defn r bal} is satisfied. 

We now check (\ref{equation: probNewDBRem}). Note that for any bit $W_i$, we already have $|N_i'|=r$ as the replication factor is $r$ is true in ${\cal D}_k$. 
%Let $N_i'$ denote the set of nodes where bit $W_i$ is stored in the new database ${\cal{D}}_k(r,[K]\setminus k)$ after the coded data rebalancing scheme for node removal . 
Now consider the event $N_i'=S$ for some $S\in \binom{[K] \setminus k}{r}$. This holds true when either of the following disjoint events happen:

\begin{enumerate}[a)]
    \item Event $E_1$: $W_i$ was stored initially at the set of nodes indexed by $S$ in the initial database itself, which implies $N_i=S$. Using the fact that the database is designed according to Lemma \ref{uniform}, we have 
    \begin{align*}
    \mathbb{P}(E_1)= \mathbb{P}(N_i=S)=\frac{1}{\binom{K}{r}}.
    \end{align*}
    \item Event $E_2$: $W_i$ was stored initially at some set of nodes indexed by $S'$ where $S' \in \binom{[K]}{r}$ such that $|S' \cap S|=r-1$ and $k \in S'$ (we call this event as $E_{21}$), and then $W_i$ was stored in node indexed by $S\setminus S'$ after rebalancing where it was part of a coded transmission by some node $\alpha\in (S\cap S')$  (we call this as event $E_{22}$). Now, $E_{12}$ means that $N_i=(S'\cap S)\cup k$ for some such $S'$, and $E_{22}$ means that the bit $W_i$ went into the box indexed by $W_{[ S\setminus S' , [K]\backslash (S\cup S') ]}^\alpha$ for some $\alpha \in (S\cap S')$. By the construction of our database, the probability of $E_{21}$ is $\frac{\binom{r}{r-1}}{\binom{K}{r}} $. By the binning technique described in Section \ref{subsection: node removal}, the probability of $E_{22}$ is $\frac{r-1}{(K-r)(r-1)}$. Hence we have,
\end{enumerate}
%Hence we have P(E_2)= P(E_21)P(E_22) =. 
%We now calculate the probabilities of the above disjoint events by applying Lemma \ref{uniform}.
%%%%%%%%%%%%%%%%%%%%%%%%%%%%%%
%%%%%%%%%%%%%%%%%%%%%%%%%%%%%&=\mathbb{P}(N_i=\{S \cap S'\} \cup k) \times \mathbb{P}(S\setminus S' \in N_i')\\
\begin{align*}
    \mathbb{P}(E_2) &= \mathbb{P}(E_{21}) \mathbb{P}(E_{22})\\  
    &= \frac{\binom{r}{r-1}}{\binom{K}{r}}\times \frac{r-1}{(K-r)(r-1)}\\
    &=\frac{r}{K-r}\frac{1}{\binom{K}{r}}.
\end{align*}
%%%%%%%%%%%%%%%%%%%%%%%%%%%%%%%%
Thus the probability that $W_i$ is stored at the set of nodes indexed by $S$ after rebalancing is given by,
\begin{align*}
%\label{newProb}
    \mathbb{P}(N_i'=S)&=\mathbb{P}(E_1)+\mathbb{P}(E_2)=\frac{1}{\binom{K}{r}}\left(\frac{K}{K-r}\right)=\frac{1}{\binom{K-1}{r}},
\end{align*}
thus proving (\ref{equation: probNewDBRem}). Using Lemma \ref{distributedDatabase}, %it follows that the expected number of bits stored in each node in the database ${\cal{D}}_k(r,[K]\setminus k)$ is given by    $\mathbb{E}(|D_n^k|)=\lambda_{rem} F, ~\forall~n \in [K]\setminus k,$ where $\lambda_{rem}=\frac{r}{K-1}$.
%This satisfies the balanced state condition of Definition \ref{defn r bal}. We 
we conclude that ${\cal{D}}(r,[K] \setminus k)$ is a decentralized $r$-balanced distributed database.
\end{IEEEproof}

In the following theorem we calculate the expected communication load of the coded data rebalancing scheme for node removal.
\begin{theorem}
\label{theorem:nodeRemoval}
Given a decentralized $r$-balanced distributed database ${\cal{D}}(r,[K])$ where a  node $k \in [K]$ is removed, the expected communication load of the coded data rebalancing scheme in Algorithm \ref{trans algorithm} for node removal is $C_{rem}=\frac{1}{r-1}$ as $F\rightarrow \infty$, and this is optimal. 
\end{theorem}
%%%%%%%%%%%%%%%
%%%%%%%%%%%%%%%%%As stated in Algorithm \ref{trans algorithm}, before XOR operation each packet involved in the transmission are padded so as to match the size of the maximum sized packet.
\begin{IEEEproof}
In order to calculate the expected communication load, we first need to know the expected size of the padded packets involved in each transmission of Algorithm \ref{trans algorithm}. Consider a transmission $X_{p_i,\boldsymbol{m}'}$ as described in Algorithm \ref{trans algorithm} sent by the node $p_i \in [K]\setminus (k \cup \boldsymbol{m}')$ where $\boldsymbol{m}' \in \binom{[K]\setminus k}{K-r-1}$. The transmission $X_{p_i,\boldsymbol{m}'}$ involves the $r-1$ packets $\{|W_{[p_l,\boldsymbol{m}']}^{p_i}|: p_l \neq p_i\}$. Before coding the packets for transmission, we pad the packets involved in $X_{p_i,\boldsymbol{m}'}$ to match the size of the largest packet among the $r-1$ packets involved. Thus the expected size of the transmission $X_{p_i,\boldsymbol{m}'}$ is given by,
\begin{align}
\label{transSize}
    \mathbb{E}(|X_{p_i,\boldsymbol{m}'}|)
    =\mathbb{E}(\max(\{|W_{[p_l,\boldsymbol{m}']}^{p_i}|: p_l \neq p_i\})).
\end{align}

We must recall that during the rebalancing operation each bit that was stored in the removed node is binned in one of the $(K-r)(r-1)$ boxes uniformly at random as shown in Figure \ref{fileSplit}. A packet $W_{[p_l,\boldsymbol{m}']}^{p_i}$ is formed by the set of bits that choose the same box labelled by $W_{[p_l,\boldsymbol{m}']}^{p_i}$. By our database construction and the binning process, the probability that a bit $W_{\tilde{i}}$ is a part of a packet $W_{[p_l,\boldsymbol{m}']}^{p_i}$ is thus given as ,
\begin{align*}
    q \triangleq \mathbb{P}(W_{\tilde{i}} \in W_{[p_l,\boldsymbol{m}']}^{p_i})=\frac{1}{\binom{K}{r}(K-r)(r-1)}, \forall {\tilde{i}} \in [F]
\end{align*}
%The size of a packet $W_{[p_l,\boldsymbol{m}']}^{p_i}$ is $l$ bits, if $l$ bits out of the $F$ bits in the database are exclusively stored in the same set of $r$ nodes indexed by $[K] \setminus \{p_l \cup \boldsymbol{m}'\}$. 
We can see that the size of a packet is a binomial random variable i.e. $|W_{[p_l,\boldsymbol{m}']}^{p_i}| \sim B(F,q)$.
Thus the probability that the size of a packet $W_{[p_l,\boldsymbol{m}']}^{p_i}$ is $l$ bits is given by,
\begin{align*}
    &\mathbb{P}(|W_{[p_l,\boldsymbol{m}']}^{p_i}|=l)\\
&=\binom{F}{l}q^l (1-q)^{F-l}. 
\end{align*}
%%%%%%%%%%%%%%%&=\mathbb{P}(\{W_{i_1},...,W_{i_l}:i_1,..,i_l \in [F]\} \in W_{[p_l,\boldsymbol{m}']}^{p_i})\\

Thus the packet sizes $\{|W_{[p_l,\boldsymbol{m}']}^{p_i}|: p_l \neq p_i\}$ are $r-1$ identically distributed binomial random variables with distribution $B(F,q).$  An asymptotic upper bound on the expected value of maximum of a finite collection of identically distributed binomial random variables is derived in Appendix \ref{expMax}. By substituting (\ref{max}) of Appendix \ref{expMax} in (\ref{transSize}), an asymptotic upper bound on the expected size of a transmission $X_{p_i,\boldsymbol{m}'}$ is given by, 
\begin{align*}
    \mathbb{E}(|X_{p_i,\boldsymbol{m}'}|)\leq F q+ \sqrt{2Fq(1-q)log(r-1)}
\end{align*}

According to Algorithm \ref{trans algorithm}, we see that $r$ transmissions are sent for every $\boldsymbol{m}'$. Let ${\cal{R}}(k,{\cal{D}},{\cal{D}}_k)$ denote our rebalancing scheme. The expected communication load of the scheme is thus given by, 
%%%%%%%%%%%%%%%%%%%for every $\boldsymbol{m}' \in \binom{[K]\setminus k}{K-r-1}$, each node $p_i \in [K]\setminus \{k \cup \boldsymbol{m}'\}$ sends a transmission $X_{p_i,\boldsymbol{m}'}$.
\begin{align*}
    &C_{rem}({\cal{R}}(k,{\cal{D}},{\cal{D}}_k))= \frac{\mathbb{E}(|X_{p_j,\boldsymbol{m}'}|) \times r \times \binom{K-1}{K-r-1}}{\lambda F}\\
    & \leq \frac{r(F q+ \sqrt{2Fq(1-q)log(r-1)})\binom{K-1}{K-r-1}}{\lambda F}
\end{align*}
Thus the expected communication load as $F\rightarrow \infty$ is given by,
\begin{align*}
    &\lim_{F\rightarrow \infty} C_{rem}({\cal{R}}(k,{\cal{D}},{\cal{D}}_k))\\
    &\leq \lim_{F \rightarrow \infty}\frac{r(F q+ \sqrt{2Fq(1-q)log(r-1)})\binom{K-1}{K-r-1}}{\lambda F}\\
    &= \frac{r\bigg(\frac{ r! (K-r)!}{K!(K-r)(r-1)}+0\bigg)\frac{(K-1)!}{r! (K-r-1)!}}{\lambda}\\
    &=\frac{1}{r-1}.
\end{align*}
  In Appendix \ref{appendix:converse}, we show that the optimal rebalancing load $C_{rem}^*(r)$ for node removal is at least $\frac{1}{r-1}$. Thus we have shown that the expected asymptotic communication load of our scheme is $\frac{1}{r-1}$ and is optimal.
\end{IEEEproof}
%This is equivalent to the case where $l$ bits out of $\{W_i,i \in [F]\}$ belong to $W_{[p_i,\boldsymbol{m}']}$ where $p_i \cup \boldsymbol{m}' \in \binom{[K]}{K-r}$. 
%The size of $W_{[p_j,\boldsymbol{m}']}$ is a binomial random variable 
%Each transmission consists of $r-1$ packets that are padded to match the maximum sized packet involved in the transmission. We require to find the expected maximum size of $r-1$ binomially distributed random variables. From Appendix \ref{expMax}, asymptotically $\mathbb{E}(max(r-1$ binomially distributed random variables$))= F q+ \sqrt{2Fq(1-q)log(r-1)}$. 
%A $\mathbb{E}($number of transmission bits sent by node $p_i)= \frac{1}{r-1} (F q+ \sqrt{2Fq(1-q)log(r-1)})$. Each of the $r$ nodes in a group send transmission bits as calculated above and there are $\binom{K-1}{K-r-1}$ groups. Hence, the communication cost, $C=\frac{r}{r-1}(F q+ \sqrt{2Fq(1-q)log(r-1)})\binom{K-1}{K-r-1}$. The expected communication load for node removal, $C_{rem}=\frac{C}{\lambda F} \simeq \frac{1}{r-1}$. 
%Under coded data rebalancing scheme for node removal, each coded transmission sent by a node consists of $r-1$ mini-packets each of which is padded to match the size of the maximum sized mini-packet. To calculate the communication load, we first need to know the expected size of mini-packet $W_{[p_j,\boldsymbol{m}']}'$ involved in each coded transmission. 
%%%%%%%%%%%%%%%%%%%%%%%%%%%5

%%%%%%%%%%%%%%%%%%%%%%%%%%%%
\begin{example}
\label{nr_eg}
\textbf{Initialisation:} Consider a system with $K=6$ nodes with replication factor $r=3$ designed as in Lemma \ref{distributedDatabase}. Each bit is stored at a set of $3$ nodes chosen uniformly at random from the set of $6$ nodes. 
This ensures that the replication factor of every bit is $3$. 
The storage content of each node consists of collections of bits that are labelled by the set of nodes in which they are not stored. For instance at node $6$, the collection of bits indexed by $W_{123},W_{124},W_{125},W_{134},W_{135},W_{145},W_{234},W_{235},W_{245}$ and $W_{345}$ will be stored, where $W_{123}$ is a convenient notation for the set of bits $W_{\{1,2,3\}}$.
% Due to random placement, we must note that all packets will not be of same size . 
%Let each bit $W_i: i \in [F]$ select a set of $r$ nodes uniformly at random from the set of $K$ nodes. $W_i$ 

\textbf{Rebalancing for node removal: } Let node $6$ be removed from the system. The replication factor of the collection of bits that were stored in node $6$ will be reduced to $2$. To restore the replication factor we perform the coded data rebalancing scheme for node removal. According to the scheme, we allow the bits in each collection of bits stored in the removed node to choose a box from a set of $3\times 2=6$ boxes. For example, each bit from the collection of bits indexed by $W_{123}$ is allowed to choose a box from a set of boxes labelled by $W_{[1,23]}^4,W_{[1,23]}^5,W_{[2,13]}^4,W_{[2,13]}^5,W_{[3,12]}^4$ and $W_{[3,12]}^5$ (where $W_{[1,23]}^4$ is a convenient notation for $W_{[1,\{2,3\}]}^4$ as given in Section \ref{subsection: node removal}). 
The bits that choose the same box are collectively called as packet and they are indexed by the label of the box they have chosen. 
A packet indexed by $W_{[1,23]}^5$, is not available at the nodes $1,2,3$ and available at nodes $4,5$. We aim to store the packet $W_{[1,23]}^5$ at node $1$ as per Algorithm \ref{trans algorithm}.

According to Algorithm \ref{trans algorithm} we perform $3$ coded transmissions for every $\boldsymbol{m}' \in \binom{[6] \setminus 6}{2}$. For $\boldsymbol{m}'=\{2,3\}$, consider the packets $W_{[1,23]}^5, W_{[4,23]}^5$ stored in node $5$. Assume that the packet $W_{[1,23]}^5$ is of bigger size than the packet $W_{[4,23]}^5$. We then pad the packet $W_{[4,23]}^5$ with zeros to match the size of packet $W_{[1,23]}^5$. Node $5$ then uses these padded packets and sends a transmission given by  $X_{5,23}=W_{[1,23]}^5 \oplus W_{[4,23]}^5$. We see that Node $1$ is the only node apart from node $5$ which has the packet $W_{[4,23]}^5$ in its storage with which it can decode its demanded packet $W_{[1,23]}^5$ from the transmission $X_{5,23}$. Similarly node $4$ can decode its demanded packet $W_{[4,23]}^5$ from the transmission $X_{5,23}$ using the packet $W_{[1,23]}^5$ which is available in its storage. We must note that the bits in the packet $W_{[1,23]}^5$ which were initially stored in the nodes $4,5,6$ is now stored at nodes $1,4,5$ after rebalancing. Thus each bit which was initially stored in the removed node $6$ is precisely stored at one extra node after rebalancing. Hence the replication factor of all the bits that were stored in the removed node $6$ is restored to $3$. \end{example}
\section{Coded Data Rebalancing for node addition}
\label{section:nodeAdditionRebalancing}
When a new empty node $K+1$ is added to the decentralized $r$-balanced distributed database ${\cal{D}}(r,[K])$, although the replication factor condition is unaffected, the balanced state condition of the database no longer holds. 
%This is because the expected number of bits stored in the new node $K+1$ is $0$. 
To restore the balanced state condition of the database we perform a rebalancing operation. After the rebalancing operation, we target to accomplish a decentralized $r$-balanced distributed database ${\cal{D}}^*(r,[K+1])=\{D_n^* \subset W: n \in [K+1]\}$ in the new system consisting of nodes $[K+1]$. Let $N_i^*$ denote the set of nodes where bit $W_i$ is stored in the new database ${\cal{D}}^*(r,[K+1])$. We then design the rebalancing scheme so that in the new database ${\cal{D}}^*(r,[K+1])$, we have
\begin{align}
\label{equation: nodeAddProb}
    \mathbb{P}(N_i^*=S)=\frac{1}{\binom{K+1}{r}},~~~~ \forall i\in[F],~~~~ \forall S \in \binom{[K+1]}{r}.
\end{align}
If the above condition holds in our new database, then we can use Lemma \ref{uniform} to show that
%,
% \begin{align*}
%     \mathbb{E}(|D_n^*|)=\lambda_{add}F, ~~~\forall n \in [K+1],
% \end{align*}
% where $\lambda_{add}=\frac{r}{K+1}$ is the new storage fraction. We can then see that 
the target database is also a decentralized $r$-balanced distributed database. We show this in Lemma \ref{nodeadditionbalancedlemma}. In Theorem \ref{nodeadditiontheoremload}, we will obtain the expected communication load of this scheme and show that this is optimal. We will next discuss the coded data rebalancing scheme for node addition case. 
%%%%%%%%%%%%%%%%%%%%%%
%Consider a node indexed by $K+1$ added to the decentralized $r$-balanced database ${\cal{D}}(r,[K])$.  Due to node addition, data skew occurs in the system and the uniform distribution of bits across the nodes gets disrupted. This necessitates a rebalancing operation among the nodes that targets to achieve a decentralized $r$-balanced database ${\cal{D}}^*(r,[K+1])=\{D_i^* \subset W: i \in [K+1]\}$. According to Definition \ref{defn r bal}, we require ${\cal{D}}^*(r,[K+1])$ to have,
% \begin{align*}
%     \mathbb{P}(N_i \in \binom{[K+1]}{r}= \frac{1}{\binom{K+1}{r}}. 
% \end{align*}
% Consider a node $j \in [K+1]$,
% \begin{align*}
%     &\mathbb{P}(N_i \in \binom{[K+1]}{r}: j \in N_i)=\frac{\binom{K}{r-1}}{\binom{K+1}{r}}=\frac{r}{K+1} \triangleq \lambda_{add},\\
%     &\mathbb{E}(|D_j|)=\lambda_{add}F,
% \end{align*}
% where $\lambda_{add}$ is the new caching fraction after rebalancing for node addition. 
% We now describe the coded data rebalancing scheme for node addition case.
%%%%%%%%%%%%%%%%%%%%%
\begin{figure}
  \includegraphics[width=\linewidth,trim={10cm 9.8cm 11cm 0 0},clip]{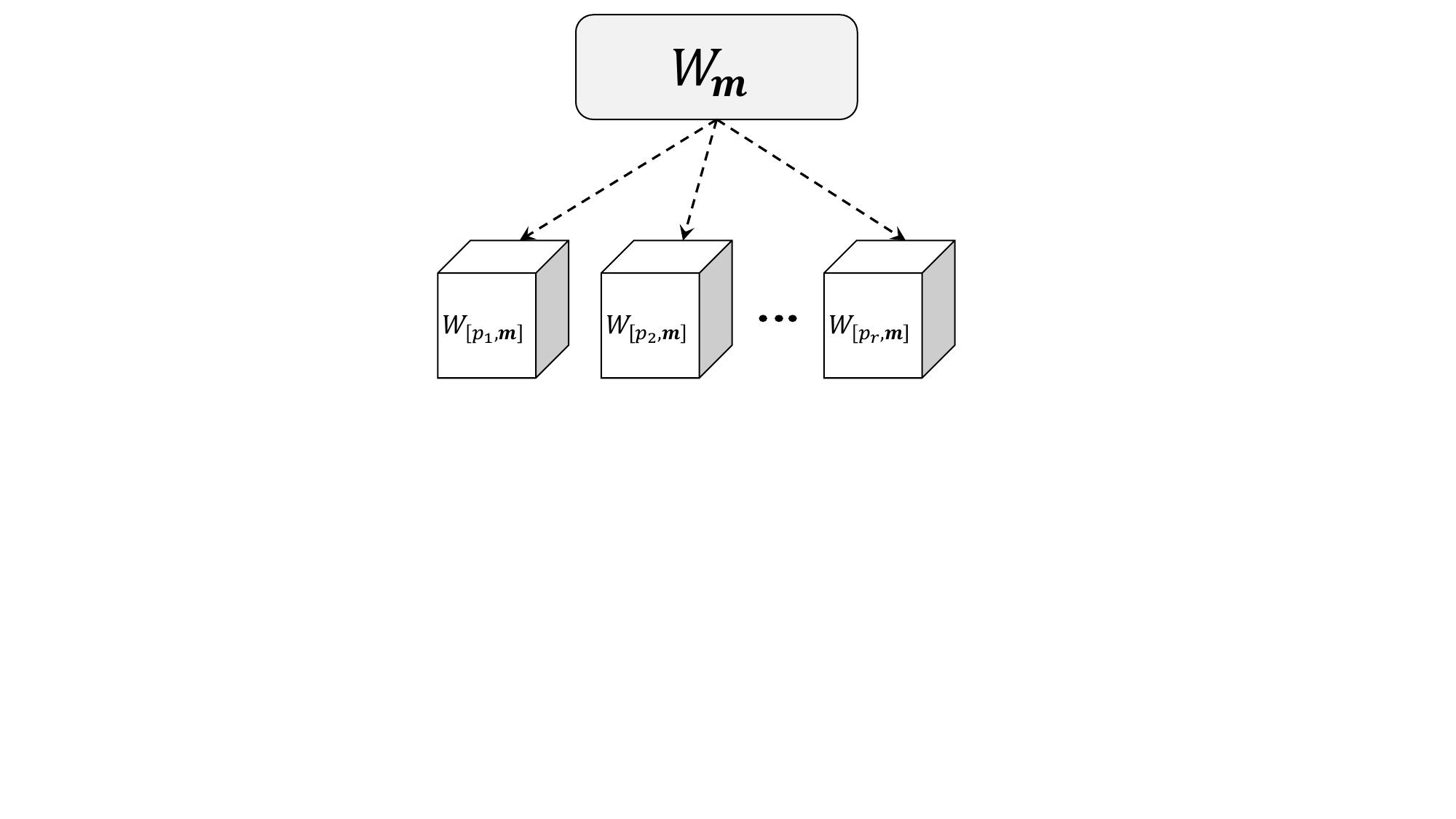}
  \caption{For every $\boldsymbol{m} \in {\cal{A}}_{[K]}$, 
  Consider $r$ boxes labelled by the set $U_{\boldsymbol{m}}$. Every bit in $W_{\boldsymbol{m}}$ is associated with one of these boxes picked with probability $\frac{1}{K+1}$, and with none of the boxes with probability $\frac{K-r+1}{K+1}$.}
  \label{nodeAdd}
\end{figure}
%%%%%%%%%%%%%%%%%%%%%%
\subsection{Coded data rebalancing scheme for node addition}

To restore the disrupted balanced state condition in the database ${\cal{D}}(r,[K])$ (designed as in Lemma \ref{distributedDatabase}) after a new node $K+1$ is added to the system, we perform the coded data rebalancing scheme for node addition. Under this scheme, each of the $K$ pre-existing nodes deletes few bits from its own storage and transmits them to the new node to establish a new decentralized $r$-balanced database ${\cal{D}}^*(r,[K+1])$. Let us index the collection of bits that were stored in the $[K]$ pre-existing nodes by ${\cal{A}}_{[K]}$ where,
\begin{align*}
    {\cal{A}}_{[K]}= \binom{[K]}{K-r}
\end{align*}

For each $\boldsymbol{m} \in {\cal{A}}_{[K]}$, we consider $r$ boxes  with labels given by the set $U_{\boldsymbol{m}}$ as follows
\begin{align}
\label{packetLabel1}
    U_{\boldsymbol{m}}=\{W_{[k,\boldsymbol{m}]}: \forall k \in [K] \setminus \boldsymbol{m}\},
\end{align}
% \begin{equation}
% \label{packetLabel2}
%     \small V_{\boldsymbol{m}}=\{W_{[\alpha_i,\boldsymbol{m}_i']}: \forall \alpha_i \in (\boldsymbol{m} \cup \{K+1\}), \boldsymbol{m}_i'=(\boldsymbol{m} \cup \{K+1\}) \setminus \alpha_i\}
%     %\boldsymbol{m}_i'\in \binom{\boldsymbol{m} \cup \{K+1\}}{K-r} , \alpha_i \in \boldsymbol{m} \cup \{K+1\} \setminus \boldsymbol{m}_i'\bigg\}.
% \end{equation}
For each $\boldsymbol{m} \in {\cal{A}}_{[K]}$, each bit in the collection $W_{\boldsymbol{m}}$ is associated with one box chosen with probability $\frac{1}{K+1}$ from the $r$ boxes in $U_{\boldsymbol{m}}$ as depicted in Figure \ref{nodeAdd}, and with probability $\frac{K+1-r}{K+1}$ the bit is not associated with any box in $U_{\boldsymbol m}$. One of the nodes in $[K]\setminus\boldsymbol{m}$ performs this binning and communicates this to each of the other nodes in $[K]\setminus\boldsymbol{m}$ so that all of them agree on the bits that are associated with their respective bins. The bits that choose the same box are collectively called as a packet and it is indexed by the label of the box that is chosen in common.

For the bits in a packet $W_{[k,\boldsymbol{m}]}$, the nodes indexed by $[K] \setminus \boldsymbol{m}$ indicates the set of nodes where it is stored initially. Also, each pre-existing node $k \in [K]$, for every $\boldsymbol{m} : k \notin \boldsymbol{m} $, there exists a packet in its storage labelled by  $W_{[k,\boldsymbol{m}]}$. According to Algorithm \ref{add trans algo}, we make each pre-existing node $k \in [K]$ to transfer the packets, 
\begin{align*}
    W_{[k,\boldsymbol{m}]} : \forall \boldsymbol{m}\in \binom{[K] \setminus k}{K-r}
\end{align*}
to the new node $K+1$ and delete these packets from its own storage. This way, the new node $K+1$ fills its storage with the packets received from each of the pre-existing nodes. Define the resultant database to be $D^*(r,[K+1])$. In the next lemma, we show that the new database $D^*(r,[K+1])$ obtained after rebalancing is a decentralized $r$-balanced distributed database. 
%%%%%%%%%%%%%%%%%%%%%%%%
\begin{algorithm}
\caption{Coded data rebalancing transmission scheme for node addition}
\label{add trans algo}
\begin{algorithmic}[1]
\Procedure{Transmission}{}
    \For {each $k \in [K]$}
    \For{each $\boldsymbol{m} \in \binom{[K] \setminus k}{K-r}$}
    \State {Node $k$ transmits $X^*_{k,\boldsymbol{m}}= W_{[k,\boldsymbol{m}]}$ to node}
    \Statex{
    \hspace{\algorithmicindent}
    \hspace{\algorithmicindent}
    \hspace{\algorithmicindent}
    $K+1$}
    \State Node $k$ deletes $W_{[k,\boldsymbol{m}]}$ from its storage
    \EndFor
    \EndFor
\EndProcedure
\end{algorithmic}
\end{algorithm}

\begin{lemma}
\label{nodeadditionbalancedlemma}
The database ${\cal{D}}^*(r,[K+1])$ is a decentralized $r$-balanced distributed database. 
\end{lemma}

\begin{IEEEproof}
At each node we check the replication factor of the bits in the packets that are transmitted to the new node. The other bits which are not part of the transmissions remain in the same $r$ nodes as in the original database, and hence their replication factor stays as $r$. 
Under Algorithm \ref{add trans algo}, for every $\boldsymbol{m} \in {\cal A}_{[K]}$ , each node $k\in [K] \backslash m$ transmits the packet 
%each pre-existing node $k \in [K]$, for every $\boldsymbol{m} \in {\cal{A}}_{[K]}$ transmits one of the packet
$W_{[k,\boldsymbol{m}]}$ to the new node $K+1$ and deletes it from its own storage. The packet $W_{[k,\boldsymbol{m}]}$ will now be stored at the $r$ nodes indexed by $[K+1]\setminus (k \cup \boldsymbol{m})$. 
%The set of packets stored at node $k$ labelled by the set $\{W_{[k,\boldsymbol{m}]}\}$ such that $k\neq k$ will be stored at the $r$ nodes indexed by $[K]\setminus (k \cup \boldsymbol{m})$. The remaining set of packets stored in node $K$ labelled by  $\{W_{[\alpha_i,\boldsymbol{m}_i']}\}$ is stored at the $r$ nodes indexed by $[K+1] \setminus \{\alpha_i \cup \boldsymbol{m}_i'\}$. Similarly we can apply the same arguments for the packets stored in the other nodes. 
%Thus, $r_i([K+1])=r$. 
This satisfies the replication factor condition of Definition \ref{defn r bal}.

We now check for the balanced state condition. Let $N_i^*$ denote the set of nodes where bit $W_i$ is stored in the new database ${\cal{D}}^*(r,[K+1])$ after performing the coded data rebalancing scheme for node addition. Consider the event $N_i^*=S \in \binom{[K+1]}{r}$. We deal with this event under the following cases :
\begin{caseof}
    \case{$K+1 \notin S$}{This case happens when $W_i$ is stored initially at a set of nodes indexed by $S$ and not transmitted to the new node by any of the nodes in $S$ during rebalancing. This event happens when $W_i$ is not associated with any box in $U_{\boldsymbol m}$, where $\boldsymbol{m}=[K]\backslash S$, which we refer to as the event $W_i\notin U_{\boldsymbol m}$. Thus probability that $W_i$ is stored in $S$ such that $K+1 \notin S$ after rebalancing is given by,
%%%%%%%%%%%%%%: K+1 \notin S
    \begin{align*}
       \mathbb{P}(N_i^*=S) &= \mathbb{P}(N_i=S) \mathbb{P}(W_i \notin U_{\boldsymbol{m}})\\
       &=\frac{1}{\binom{K}{r}}.\frac{K-r+1}{K+1}\\
       &=\frac{1}{\binom{K+1}{r}}
    \end{align*}
    }
    \case{$K+1 \in S$}{This event happens when both of the following events happen. 
\begin{itemize}
\item $W_i$ has been initially stored in some set of nodes indexed by $S'$, where $S' \in \binom{[K]}{r}$ such that $|S\cap S'| = r-1$. We call this event as $E_1$. As there are $K-(r-1)$ such choices of $S'$ for a chosen $S$,  $E_1$ happens with probability $\frac{(K-(r-1))}{\binom{K}{r}}$ (by design of our initial database).  
\item $W_i$ is then transmitted by the node indexed by $S' \setminus S$ to the new node $K+1$ and then deleted from the storage of node $S' \setminus S$. This event occurs when $W_i$ is a part of a packet labelled by $W_{[k,\boldsymbol{m}]}$, as defined in (\ref{packetLabel1}), where $\boldsymbol{m}=[K]\backslash S'$, and $k = S' \setminus S$. We call this event $E_2$. This event happens with probability, $\frac{1}{K+1}$ (as there are $K+1$ boxes which this bit can go into as per our description in this section). 
\end{itemize} 

Thus, the probability that $N_i^*=S$ where $K+1 \in S$ is given by,
%P(N_i^*=S)=P(E_1)P(E_2)= 
    \begin{align*}
        \mathbb{P}(N_i^*=S) &= \mathbb{P}(E_1) \mathbb{P}(E_2)\\
        &=\frac{K-(r-1)}{\binom{K}{r}}\frac{1}{K+1}\\
        &=\frac{1}{\binom{K+1}{r}}
    \end{align*}
    }
\end{caseof}
%%%%%%%%%%%%%%
%This case happens when $W_i$ is stored initially at a set of nodes indexed by $S'$ where $S' \in \binom{[K]}{r}$ such that $|S \cap S'|=r-1$. $W_i$ is then transmitted by the node indexed by $S \setminus S'$ to the new node $K+1$ and then deleted from the storage of node $S \setminus S'$. This event occurs when $W_i$ is a part of a packet labelled by the set $W_{k,\boldsymbol{m}}$ as defined in (\ref{packetLabel1}) such that $k = S \setminus S'$. The probability of the event where $W_i$ is stored in $S: K+1 \in S$ after rebalancing is given by,
% Based on the above events we now calculate the probability of 
% Case (i): 
% \begin{align*}
%     &\mathbb{P}(N_{i,add} \in \binom{[K+1]}{r}: K+1 \in N_{i,add})\\
%     &=\mathbb{P}(N_i \in \binom{[K]}{r}: N_{i,add} \cap N_i=r-1)\\
%     & \times \mathbb{P}(W_i \in W_{[k,[K]\setminus N_{i,add}]}: k=N_{i,add}\setminus N_i)\\
%     &=\frac{1}{\binom{K}{r}}(K-r+1) \times \frac{1}{K+1}\\
%     &=\frac{1}{\binom{K+1}{r}}.
% \end{align*}
% Case (ii):  
% \begin{align*}
%     &\mathbb{P}(N_{i,add} \in \binom{[K+1]}{r} : K+1 \notin N_{i,add})\\
%     &=\mathbb{P}(N_i \in \binom{[K]}{r}) \times \mathbb{P}(W_i \in W_{[k,\boldsymbol{m}]} : K+1 \in \{k \cup \boldsymbol{m}\})\\
%     &=\frac{1}{\binom{K}{r}}\times\frac{K-r+1}{K+1}\\
%     &=\frac{1}{\binom{K+1}{r}}.
% \end{align*}
By the above probability distribution, and using Lemma \ref{uniform}, we can see that 
%
%the expected number of bits stored in each node in the database ${\cal{D}}^*(r,[K+1])$ is given by,
% \begin{align*}
%     \mathbb{E}(|D_n^*|)=\lambda_{add}F, \forall n \in [K+1],
% \end{align*}
% where $\lambda_{add}=\frac{r}{K+1}$ is the new storage fraction. Thus the balanced state condition of Definition \ref{defn r bal} is satisfied. Hence, 
${\cal{D}}^*(r,[K+1])$ is a decentralized $r$-balanced distributed database.
\end{IEEEproof}
%%%%%%%%%%%%%%%%%%%%%%
%In both cases we see that bit $W_i$ is randomly and uniformly distributed among the $K+1$ nodes. 

% Consider a node $j \in [K+1],$
% \begin{align*}
%     &\mathbb{P}(N_{i,add} \in \binom{[K+1}{r}: j \in N_{i,add})\\
%     &=\frac{\binom{K}{r-1}}{\binom{K+1}{r}}=\frac{r}{K+1}=\lambda_{add},\\ 
% &\mathbb{E}(|D_j|)=\lambda_{add} F.
% \end{align*}
% Hence, we have obtained a decentralized $r$-balanced database ${\cal{D}}(r,[K+1])$.

In the following theorem we calculate the expected communication load of the coded data rebalancing scheme for node addition.

%%%%%%%%%%%%%%%
\begin{theorem}
\label{nodeadditiontheoremload}
Given a decentralized $r$-balanced distributed database ${\cal{D}}(r,[K])$ where a new node $K+1$ is added, the expected communication load of the coded data rebalancing scheme for node addition given in Algorithm \ref{add trans algo}  is $C_{add}=1$, and this is optimal.
\end{theorem}
%%%%%%%%%%%%%%%
\begin{IEEEproof} 
We first need to calculate the expected size of the packet involved in each transmission. Consider a transmission $X^*_{k, \boldsymbol{m}}$ as described in Algorithm \ref{add trans algo} sent by the node $k \in [K] \setminus \boldsymbol{m}$ where $\boldsymbol{m} \in \binom{[K]}{K-r}$. The transmission $X^*_{k, \boldsymbol{m}}$ consists of the packet $W_{[k, \boldsymbol{m}]}$ which is transmitted to the new node $K+1$.
% The expected size of the transmission $X^*_{k, \boldsymbol{m}}$ sent by node $k$  is given by, 
% \begin{align}
% \label{trans size}
%     \mathbb{E}(|X^*_{k, \boldsymbol{m}}|)=\mathbb{E}(|W_{[k, \boldsymbol{m}]}|)
% \end{align}

%During the rebalancing operation we allowed each bit in the collection of bits stored in the pre-existing nodes to choose one of the $(K+1)$ boxes uniformly at random as illustrated in Figure \ref{nodeAdd}. We know that the bits that choose the same box labelled by $W_{[k, \boldsymbol{m}]}$ are collectively called as packet indexed by $W_{[k, \boldsymbol{m}]}$.

By the design of our initial database and by the binning of the bits in $W_{\boldsymbol{m}}$, the probability that $W_i$ is a part of the packet $W_{[k,\boldsymbol{m}]}$ is given by,
%Let the probability that a bit $W_i$ is a part of the packet $W_{[k, \boldsymbol{m}]}$ be defined by, 
\begin{align*}
    q' \triangleq \mathbb{P}(W_i \in W_{[k, \boldsymbol{m}]}) =\frac{1}{\binom{K}{K-r}(K+1)}, \forall i \in [F].
\end{align*}
 
%The size of a packet $W_{[k, \boldsymbol{m}]}$ is $l$ bits, if $l$ bits out of the $F$ bits in the database are exclusively stored in the same set of $r$ nodes indexed by $[K] \setminus \boldsymbol{m}$. 
We can see that the size of a packet is a binomial random variable which implies $|W_{[k, \boldsymbol{m}]}| \sim B(F,q')$. 
Thus the probability that the size of a packet $W_{[k, \boldsymbol{m}]}$ is $l$ bits is given by,
\begin{align*}
    &\mathbb{P}(|W_{[k, \boldsymbol{m}]}|=l)\\
    %&= \mathbb{P}(\{W_{i_1},...,W_{i_l}: i_1,...,i_l \in [F]\} \in W_{[k, \boldsymbol{m}]})\\
    &=\binom{F}{l}(q')^l(1-q')^{F-l}.
\end{align*}
Hence, the expected size of a transmission $X_{k, \boldsymbol{m}}^*$ is given by, 
\begin{align*}
    \mathbb{E}(|X^*_{k, \boldsymbol{m}}|)=\mathbb{E}(|W_{[k, \boldsymbol{m}]}|)= Fq'
\end{align*}

According to Algorithm \ref{add trans algo}, each node $k \in [K]$ sends a transmission $X_{k, \boldsymbol{m}}^*$ to the new node $K+1$ for every $\boldsymbol{m} \in \binom{[K]\setminus k}{K-r}$. There are $\binom{K-1}{K-r}$ transmissions sent by every node $k \in [K]$. Hence the expected communication load is given by,
\begin{align*}
    C_{add}&=\frac{\mathbb{E}(|X^*_{k, \boldsymbol{m}}|) \times K \binom{K-1}{K-r}}{\lambda_{add}F}\\
    &=\frac{F \times K  \binom{K-1}{K-r}}{\binom{K}{K-r}(K+1)\lambda_{add}F}\\
    &=\frac{Fr}{(K+1)\lambda_{add}F}=1
\end{align*}
Therefore we have obtained the expected communication load of our rebalancing scheme. A matching lower bound is easily obtained by a cut-set argument, seeing that the new node is empty when it is added to the system. Hence, our expected load is optimal. 
\end{IEEEproof}
% The coded data rebalancing scheme for node addition allows each pre-existing node $k \in [K]$ to transmit sub-packets to the new node $[K+1]$ and deletes them from the $k^{th}$ node's cache. Each packet $W_{\boldsymbol{m}}: \boldsymbol{m} \in \binom{[K]\setminus k}{K-r}$ is split into $K+1$ equal sized sub-packets. 
% %The size of the sub-packet $W_{[k,\boldsymbol{m}]}$ is $\frac{1}{K+1}$ times the size of the packet $W_{\boldsymbol{m}}$. 
% Assuming uniform probability, let $\mathbb{P}(W_i \in W_{\boldsymbol{m}})=\frac{1}{\binom{K}{K-r}}=q$.
% The size of a packet $W_{\boldsymbol{m}}$ is $l$ bits, if $l$ bits out of the $F$ bits in the database are exclusively cached in the same set of $r$ nodes and not cached in the remaining nodes indexed by $\boldsymbol{m} \in \binom{[K]}{K-r}$. 
% \begin{align*}
%     &\mathbb{P}(|W_{\boldsymbol{m}}|=l: l \in \{0,1,..,F\})\\
% &=\mathbb{P}(\{W_{i_1},...,W_{i_l}\} \in W_{\boldsymbol{m}}: \boldsymbol{m} \in \binom{[K]}{K-r})\\
% &=\binom{F}{l}q^l (1-q)^{F-l}\\
% \end{align*}
% \[\mathbb{E}(|W_{\boldsymbol{m}}|)=Fq\]
% \[\mathbb{E}(|W_{[k,\boldsymbol{m}]}|)=\frac{Fq}{K+1}=\frac{F}{\binom{K}{K-r}(K+1)}.\]

% The number of sub-packets transferred by any node $k \in [K]$ to the new node $K+1$ is $\binom{K-1}{K-r}$. The communication cost, \[C=K\times \binom{K-1}{K-r} \times \frac{F}{\binom{K}{K-r}(K+1)}=\frac{Fr}{K+1}=\lambda_{add} F\].
% Hence, the expected communication load for node addition, $C_{add}=1$.

%%%%%%%%%%%%%%%%%%%%%

\begin{example}
\label{na_eg}
\textbf{Initialisation:} Consider a system with $K=4$ nodes with replication factor $r=2$ designed as per Lemma \ref{distributedDatabase}. Each bit is stored at a set of $2$ nodes uniformly at random from the set of $4$ nodes. This ensures that the replication factor of every bit is $2$. The collection of bits stored at every node is indexed by the set of nodes where it is not stored. For instance, at node indexed by $1$, the collection of bits indexed by $W_{23},W_{24}$ and $W_{34}$ will be stored. 

\textbf{Rebalancing for node addition: } Let a new node indexed by $5$ be added to the system. Since the new node arrives without any data in its storage, the expected number of bits stored in the new node is $0$. To restore the balanced state condition in the database we perform the coded data rebalancing scheme for node addition. According to the scheme, each bit in the collection of bits stored across the pre-existing nodes is allowed to choose a box from a set of $r=2$ boxes chosen with probability $\frac{1}{K+1}=\frac{1}{5}$, or choose to be not in any of the 2 boxes with probability $\frac{3}{5}$. For example, each bit from the collection of bits indexed by $W_{23}$ is allowed to choose a box with probability $\frac{1}{5}$ from a set of boxes labelled by $W_{[1,23]}$ and $W_{[4,23]}$. The bits that choose the same box are called as packet and they are indexed by the label of the box they have chosen in common.

We perform transmissions according to Algorithm \ref{add trans algo}. If we consider node $1$, for $\boldsymbol{m}=\{2,3\}, \{2,4\}, \{3,4\}$ it transmits packets indexed by $W_{1,23}, W_{1,24}$ and $W_{1,34}$ respectively to the new node and deletes them from its own storage. Similarly every pre-existing node $k \in [4]$ sends $3$ transmissions. Thus, the new node $4$ fills its storage with the transmissions received from the pre-existing nodes. 

% We first split each packet in the database into $K+1$ equal sized sub-packets. Packet $W_{234}$ cached in node $1$ will be split into sub-packets $W_{[1,234]},W_{[5,234]},W_{[6,234]},W_{[2,346]},W_{[3,246]},W_{[4,236]}$. Node $1$ transmits sub-packet $W_{[1,234]}$ to the new node $6$ and deletes it from its cache. New node $6$ receives such sub-packets from each of the $5$ pre-existing nodes and stores them in its cache.  
\end{example}

%%%%%%%%%%%%%%%%%%%%%%%%
\begin{appendices}
\section{Converse for node removal}
\label{appendix:converse}
This converse essentially follows from similar arguments as in \cite{codedData} (proof of Theorem 1 in \cite{codedData}). However for the sake of completeness we give the complete proof here. This proof also uses some simpler alternate arguments than that in \cite{codedData}. This proof, as in \cite{codedData}, uses the induction based technique developed in \cite{FundLimitsDistribCom}. 

%We first give the arguments for the node removal case. 
%%%%%%%%%%%%%%%%%%%%%
Let each of the $F$ bits of the data be chosen independently and uniformly at random from $\{0,1\}$ . Without loss of generality, we assume that node $K$ has been removed from the system. Thus, each bit of $D_K$ (storage in $K$) has to be placed in exactly one surviving node. As in Section \ref{sys_model}, let $P_k:k\in[K-1]$ denote the set of bits of $D_K$ to be placed in nodes $k\in[K-1]$ respectively. We also have $P_k\cap P_{k'}=\phi, \forall k\neq k'.$

For some $k,S\subset [K]$, we define the the quantity 
$a_k^S$ as the number of bits of $P_k$ which are available exclusively in at least one node of $S$, i.e., 
%%%%
\begin{align*}
a_k^S=|P_k\bigcap\left(\bigcup_{k_1\in S}D_{k_1}\right)\backslash\left(\bigcup_{k_2\in [K-1]\backslash S}D_{k_2}\right)|.
\end{align*}
%%%
Let $X_k=\phi_k(D_k)$ denote the transmission by node $k$ in a valid rebalancing scheme. For $S\subset[K-1]$, we define $$X_S\triangleq \{X_k:k\in S\}.$$ Let $$Y_S=\bigcup_{k\in S}P_k\cup D_k.$$

We want to show that 
\begin{align}
\label{eqn401}
H(X_{[K-1]})\geq \frac{|D_K|}{r-1}.
\end{align}
The converse to Theorem \ref{theorem:nodeRemoval} will then follow by noting that $C_{rem}\geq \frac{{\mathbb E}(H(X_{[K-1]}))}{\lambda F}$.

We shall use the following claim to show our proof of (\ref{eqn401}). 
%%%
\begin{claim}
\label{claim1}
\begin{align*} 
H(X_S|Y_{\overline S})\geq \frac{\sum\limits_{k\in S}a_k^S}{r-1},
\end{align*}
where $\overline{S}=[K-1]\setminus S.$
\end{claim}
%%%
Once we show Claim \ref{claim1}, we can plug $S=[K-1]$, and (\ref{eqn401}) follows as $\sum_{k\in [K-1]}a_k^{[K-1]}=|D_K|.$ We now prove Claim \ref{claim1} through induction on $|S|$. 

Consider the base case that $|S|=2$, and let $S=\{1,2\}$ without loss of generality. We know that $r\geq 2$. If $r>2,$ then $a_1^{\{1,2\}} = a_2^{\{1,2\}} = 0$, as the bits demanded by node $k \in \{1,2\}$ are available in at least $2$ other survivor nodes apart from $k$. Thus the claim is verified in this subcase. Now, suppose that $r=2$. Then clearly $H(X_{\{1,2\}}|Y_{\overline S})\geq a_1^{\{2\}}+a_2^{\{1\}},$ as the bits demanded by node $2$ and present exclusively in node $1$ (and in no other node) has to be sent to node $2,$ and vice-versa. As the bits demanded by node $i$ are not available at node $i,$ $a_1^{\{1,2\}}=a_1^{\{2\}},$ and similarly $a_2^{\{1,2\}}=a_2^{\{1\}}.$ Thus we have $H(X_{\{1,2\}}|Y_{\overline S})\geq a_1^{\{1,2\}}+a_2^{\{1,2\}}$. Hence, the claim is satisfied for this subcase also. This completes the base case $|S|=2$.

Now we assume that the claim holds for $|S|=t$ and show that it is true for $|S|=t+1.$
\begin{align*}
    H(X_S|Y_{\overline{S}})& =\frac{1}{|S|} \sum\limits_{k \in S} H(X_S,X_k|Y_{\overline{S}})\\
    & \geq \frac{1}{|S|}  \Bigg(\sum\limits_{k \in S} H(X_S|X_k,Y_{\overline{S}})+H(X_S|Y_{\overline{S}}) \Bigg)
\end{align*}
By reordering the terms we get,
\begin{equation}
\label{sub}
\begin{split}
    H(X_S|Y_{\overline{S}}) & \geq \frac{1}{t} \sum\limits_{k \in S} H(X_{S\backslash k}|X_k,Y_{\overline{S}})\\
    & \geq \frac{1}{t} \sum\limits_{k \in S} H(X_{S\backslash k}|X_k,D_k,Y_{\overline{S}})\\
    & = \frac{1}{t} \sum\limits_{k \in S} H(X_{S\backslash k}|D_k,Y_{\overline{S}}),
\end{split}
\end{equation}
where the last equality follows as $H(X_k|D_k)=0$.

Now, we have that $0=H(P_k|X_{S\backslash k},D_k,X_{\overline S})\geq H(P_k|X_{S\backslash k},D_k,\{D_{k_1}:k_1\in {\overline S}\})\geq H(P_k|X_{S\backslash k},D_k,Y_{\overline S}).$ Thus, $H(P_k|X_{S\backslash k},D_k,Y_{\overline S})=0.$ Using this in the above equation, we get
%%%%
\begin{align}
\nonumber
H(X_S&|Y_{\overline{S}}) \geq \frac{1}{t} \sum\limits_{k \in S}H(X_{S\backslash k},P_k|D_k,Y_{\overline{S}})\\
\nonumber
&=\frac{1}{t}\sum\limits_{k \in S}\left(H(P_k|D_k,Y_{\overline{S}})+H(X_{S\backslash k}|P_k,D_k,Y_{\overline{S}})\right)\\
\label{eqnAks}
&=\frac{1}{t}\left(\sum\limits_{k \in S}a_k^{S}+\sum\limits_{k \in S}H(X_{S\backslash k}|Y_{\overline{S\backslash k}})\right)\\
\label{eqn411}
&\geq \frac{1}{t}\left(\sum\limits_{k \in S}a_{k}^{S}+\frac{\sum\limits_{k\in S}\sum\limits_{k_1\in S\backslash k}a_{k_1}^{S\backslash k}}{r-1}\right)\\
\label{eqn412}
&\geq \frac{1}{t}\left(\sum\limits_{k \in S}a_{k}^{S}+\frac{\sum\limits_{k_1\in S}\sum\limits_{k\in S\backslash k_1}a_{k_1}^{S\backslash k}}{r-1}\right),
\end{align}
%%%%
where (\ref{eqnAks}) follows as $H(P_k|D_k,Y_{\overline S})=a_k^S$ and (\ref{eqn411}) follows by the induction hypothesis. Now the term $a_{k_1}^{S\setminus k_1}$ consists of all the bits demanded by $k_1$ available in some collection of precisely $r-1$ nodes of $S$ exclusively and not in $\overline{S}$. Note that $a_{k_1}^{S}=a_{k_1}^{S\backslash k_1}$. Consider any bit $b$ demanded by $k_1$ present exclusively in $r-1$ nodes of $S\backslash k_1$, which we refer to as $S_b\subseteq S\setminus k_1$. For any $k\in S\backslash k_1$, let $P_{k_1}^{S\backslash k}$ denote the set of bits demanded by $k_1$ and present in exclusively $r-1$ nodes of $S\backslash k.$ Then $b\in P_{k_1}^{S\backslash k}$ if and only if $S_b\subset {S\backslash k}.$ This happens whenever $k\in {S\backslash (S_b\cup k_1)}$, i.e., exactly $|S\backslash k_1|-|S_b|=(t-(r-1))$ times as we run through all $k\in S\backslash k_1$. Thus we have that $$\sum\limits_{k\in S\backslash k_1}a_{k_1}^{S\backslash k}=(t-r+1)a_{k_1}^{S}.$$ Plugging this in (\ref{eqn412}) completes the proof of the claim and hence proves (\ref{eqn401}), and thus the converse for node removal is complete. 
%%%%
\section{Expectation of maximum of Binomially distributed random variables}
\label{expMax}
Let $X=\{X_1,X_2,...,X_r\}$ be a set of $r$ identically distributed binomial random variables such that $X_i \sim B(n,p)$. We intend to calculate an upper bound on the expected value of the maximum of $X_i$s. From Jensen's equality we know for positive $t$,  
\begin{align*}
    \exp(t~ \mathbb{E}(\underset{i}{\max}~ X_i))&\leq \mathbb{E}(\exp(t~ \underset{i}{\max}~ X_i))\\
    &=\mathbb{E}(\underset{i}{\max} \exp(tX_i))
\end{align*}
For large $n$ and constant $p$, using the central limit theorem, the binomial distributed $B(n,p)$ can be approximated by the Gaussian distribution ${\C{N}}(\mu,\sigma^2)$ where $\mu = np$ and $\sigma^2= np(1-p)$. Using this, the fact that the $\max\limits_i \exp(tX_i)\leq \sum\limits_i \exp(tX_i)$, and by linearity of expectation, 
\begin{align*}
    \exp(t~ \mathbb{E}(\underset{i}{\max}~ X_i))&\leq \sum\limits_{i=1}^r \mathbb{E}(\exp(tX_i))\\
    &= r \exp\bigg(t\mu+\frac{t^2 \sigma^2}{2}\bigg)
\end{align*}
Thus, we obtain 
\begin{align}
    \label{mgf}
    \mathbb{E}(\underset{i}{\max}~ X_i) &\leq \frac{\log r}{t}+\mu+\frac{t \sigma^2}{2}
\end{align}
The R.H.S of (\ref{mgf}) reaches its minimum at  $t=\frac{\sqrt{2 \log{r}}}{\sigma}$. Substituting this minimizing value of $t$ in (\ref{mgf}), we get
\begin{align}
\label{max}
    \mathbb{E}(\underset{i}{\max}~ X_i) &\leq \frac{\sigma\sqrt{\log{r}}}{\sqrt{2}}+ np+ \frac{\sigma \sqrt{2 \log r}}{2}\\
    &= np+\sqrt{np(1-p) 2 \log r}
\end{align}
\end{appendices}
\bibliographystyle{IEEEtran}
\bibliography{IEEEabrv,RandomPlacementCodedDataRebalancing.bib}
\end{document}